# Identifying the Threshold Chain Length for Stress Overshoot in Ring-Linear Polymer Blends under Uniaxial Elongation: The Role of Multiple Threading

*Takahiro Murashima,*[1] Katsumi Hagita,[2] and Toshihiro Kawakatsu[3]*

[1]Department of Physics, Tohoku University, 6-3, Aramaki-aza-Aoba, Aoba-ku, Sendai, 980-8578, Japan

[2]Department of Applied Physics, National Defense Academy, 1-10-20 Hashirimizu, Yokosuka, 239-8686, Japan

[3]Admissions Center, Tohoku University, 28, Kawauchi, Aoba-ku, Sendai, 980-8576, Japan

*murasima@cmpt.phys.tohoku.ac.jp

Graphical Abstract

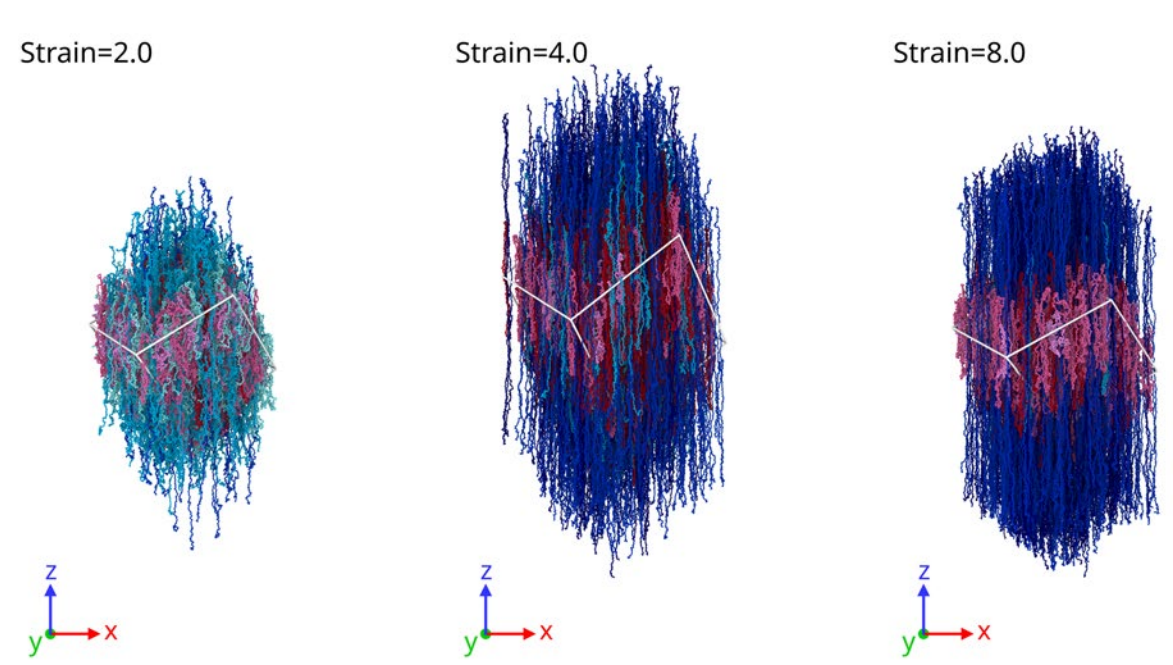

for Table of Contents use only (8.25cm vs 4.45cm)

ABSTRACT. The rheological behavior of ring-linear polymer blends under uniaxial elongational flow has remained a subject of intense debate, particularly regarding the emergence of stress overshoot. Herein, we employ coarse-grained molecular dynamics simulations to investigate the chain-length dependence of elongational viscosity in 1:1 ring-linear blends of flexible chains with the equal molecular weight. Our results reveal a distinct threshold in the degree of threading, quantified by the number of entanglements $Z = N/N_e$ (where $N$ is the number of beads per chain and $N_e$ is the entanglement chain length), for the appearance of stress overshoot: while blends with shorter chains ($Z \leq 2$) exhibit monotonic stress growth, a clear stress overshoot emerges when the chain length reaches a threshold value ($Z \approx 4$). Consistent with previous reports, this overshoot originates from a thread-to-unthread transition. At the threshold chain length, multiple linear chains penetrate a single ring, providing sufficient topological constraints to significantly stretch the ring under elongational flow. We predict that this transition can be experimentally validated via 2D small-angle neutron scattering patterns in the plane of the stretching and perpendicular directions, offering a direct structural signature of the ring recoil process for future experimental verification.

## 1. Introduction

Ring polymers have remained a long-standing enigma in polymer physics due to their unique, end-free topology.[1] A major turning point in the history of this field was the landmark study by Kapnistos et al..[2] They demonstrated that the stress relaxation of high-purity ring polymer melts follows a power-law behavior distinct from the conventional reptation theory for linear chains.[3] To achieve the necessary purity, they utilized liquid chromatography at the critical condition[4] (LCCC), which enabled the efficient separation of topologically pure rings from linear chain contaminants. They further experimentally revealed that even a trace amount of linear chain contaminants can lead to the “threading” of rings by linear chains, creating a “pinning” effect that dramatically retards the relaxation dynamics. Since this discovery, elucidating the microscopic nature of these topological constraints—specifically how threading events dictate the size and diffusivity of the rings—has been a subject of intensive research using molecular simulations,[5–10] pulsed field gradient NMR spectroscopy,[11] and, more recently, advanced neutron scattering techniques.[12,13] In parallel, numerous studies have highlighted the complex rheological consequences of such interactions, identifying a non-monotonic dependence of viscosity on blend composition where the mixture viscosity can even surpass that of the pure linear counterpart.[14–18]

In equilibrium and linear response regimes, these topological constraints cause the rings to become trapped within the transient entanglement network of the surrounding linear chains. These trapped rings can only relax through constraint release induced by the motion of the linear chains, which leads to a dramatic increase in the blend viscosity.[16] Beyond the linear regime, shear rheology of marginally entangled rings and their blends with linear chains has revealed that rings exhibit a weaker stress overshoot and milder shear thinning than their linear precursors, indicating that the rings undergo a smaller effective deformation under flow.[18] Furthermore, recent coarse-grained molecular dynamics simulations have reported that the threading between ring and linear chains forms a composite entanglement topology, which acts as an additional source of friction under non-linear flow conditions.[19] Direct evidence for such intermittent topological interactions has been provided by single-molecule observations. Zhou et al.[20,21] used fluorescence microscopy to visualize the dynamics of individual ring DNA molecules in semidilute linear DNA solutions under extensional flow. They discovered that rings exhibit strikingly large conformational fluctuations even in the steady-state regime, which they attributed to the transient threading and unthreading of linear chains through the stretched rings.

These molecular-scale observations strongly support the idea that the coupling between ring and linear components is governed by a dynamic topological process.

The most singular phenomenon in uniaxial elongational flow is the emergence of a stress overshoot in the transient elongational viscosity. Borger et al.[22] identified that this overshoot is driven by a "thread-to-unthread" transition. Immediately after the start-up of uniaxial elongation, the threaded linear chains act as moving pulleys, forcibly stretching the ring chains and causing an accumulation of excessive tension within them. Subsequently, as the strain increases, the linear chains unthread from the rings, leading to a sudden release of the accumulated tension, which is observed macroscopically as a stress drop (overshoot).

However, it has been suggested that the presence or absence of this overshoot phenomenon strongly depends on the number of entanglements per chain, $Z$. The systems for which clear overshoots under uniaxial elongation have been reported—including the experiments and simulations of Borger et al.[22] and the large-scale simulations of O'Connor et al.[19]—all utilized sufficiently long chains with $Z \approx 14$. In contrast, we showed in our previous work[23] that for ring-linear blends of flexible chains with $Z \approx 2$, a prominent overshoot was observed in biaxial elongational flow, whereas no such overshoot was detected under uniaxial elongational flow. Since it remained unclear whether this absence was due to the inherent nature of flexible chains or simply an insufficient degree of entanglement, investigating the emergence of an overshoot in longer flexible systems is of fundamental importance. These collective results suggest that a threshold chain length is required to generate a powerful "topological grip"[23] that is sufficient to overstretch the rings and induce a macroscopic stress peak under uniaxial elongation. Furthermore, previous analyses have shown that as the size of the ring chain increases, the number of penetrating linear chains ($n_p$) also increases, which induces the expansion of the ring conformation.[5–13] This strongly suggests that if the chain length is sufficiently long, a "multiple threading" state—where multiple linear chains interact with a single ring—is formed, potentially producing a more strong stretching effect under elongation.

In this study, we perform coarse-grained molecular dynamics simulations of ring-linear homopolymer blends with the equal molecular weight. By systematically varying the chain length ($Z$ = 1, 2, and 4), we investigate the chain-length dependence of the non-linear elongational rheology. Specifically, we focus on the transition from the unentangled to the moderately entangled regime to clarify whether a critical chain length exists for the emergence of a macroscopic stress overshoot. Our analysis aims to elucidate the role of multiple threading and how the number of topological constraints dictates the transient conformational tension of the rings under strong flow conditions. By combining stress decomposition, topological analysis using the Gauss linking number, and real-space density profiles, we provide a consistent physical picture of how multiple threading dictates the non-linear signatures of ring-linear blends. Furthermore, we predict the 2D small-angle neutron scattering (SANS) patterns in the plane of the stretching and perpendicular directions associated with this transition, offering a roadmap for future experimental validation using elongational rheometers coupled with neutron beamlines.

## 2. Simulation Method

We employed the coarse-grained bead-spring model developed by Kremer and Grest[24] to simulate the dynamics of ring-linear polymer blends. In this model, all beads interact via the truncated-and-shifted Lennard–Jones (LJ) potential, which accounts for excluded volume effects. Adjacent beads along a chain are connected by the finite extensible nonlinear elastic (FENE) potential. For the flexible Kremer–Grest model (with no bending potential), previous estimates of the entanglement length have varied between $N_e$ = 65 and 85, depending on the specific topological analysis method and finite chain length effects.[25–29] Recent comprehensive studies,[30–32] however, have provided more refined estimates by accounting for long-range bond correlations induced by melt incompressibility and by employing $N \to \infty$ extrapolation to eliminate systematic errors associated with chain ends. Following these advancements, which identify $N_e \approx 80$ as the asymptotic limit for properly equilibrated, long-chain flexible melts, we adopt $N_e$ = 80 as the characteristic entanglement length in this study. We investigated equi-weight (1:1 ratio) blends of ring and linear polymers with identical chain lengths of $N$ = 80, 160, and 320, which correspond to $Z$ = 1, 2, and 4, respectively. The choice of the equi-weight composition is based on the findings of O'Connor et al.,[19] who explored various ring-linear mixing ratios and reported that the magnitude of the stress overshoot is maximized in the 1:1 blend. This composition is also of particular experimental interest; Kruteva et al.[13] recently demonstrated that the zero-shear viscosity of symmetric ring‑linear blends reaches a maximum at the 1:1 blend, where the ring dynamics undergo a distinct crossover from self-similar relaxation to a regime governed by the transient entanglement network of the linear matrix. By selecting this ratio, we aim to capture the most prominent rheological signatures of the topological interactions between the two components. The total number of particles in each simulation was $N_{tot}$ = 614,400 for all chain lengths considered ($N$ = 80, 160, and 320). For comparison, pure linear and pure ring melts were also simulated as baseline systems.

All physical quantities are reported in standard LJ units, where the bead diameter $\sigma$, the depth of the LJ potential well $\varepsilon_{LJ}$, and the bead mass $m$ are set to unity. The characteristic time scale is defined as $\tau = \sigma\sqrt{\frac{m}{\varepsilon_{LJ}}}$. The system temperature was maintained at $T = 1.0\frac{\varepsilon_{LJ}}{k_B}$. For both the equilibrium and non-equilibrium molecular dynamics (NEMD) simulations under uniaxial elongation, we consistently employed Langevin dynamics with a damping parameter of $2.0\tau$, corresponding to a friction coefficient of $\gamma = 0.5\tau^{-1}$, to maintain a stable isothermal condition. The monomer density was set to $\rho = 0.85\sigma^{-3}$. In the FENE potential, for bonded interactions, the spring force constant $k = 30\frac{\varepsilon_{LJ}}{\sigma^2}$ and the maximum bond extension length $R_0 = 1.5\sigma$. For non-bonded interactions, the truncated-and-shifted LJ potential with a cut-off distance of $r_{\mathrm{cut}} = 2^{1/6}\sigma$ acts as a purely repulsive potential. Given the number density $\rho$, the volume of the simulation box is $V = N_{\mathrm{tot}} / \rho$, resulting in a cubic box length of $L = V^{1/3} \approx 89.8\sigma$. This system size was chosen to ensure that the cubic simulation box length is well exceeds not only the radius of gyration ($R_{\mathrm{g}} = 9.4\sigma$) but also the mean end-to-end distance ($R_{\mathrm{ee}} = 22.1\sigma$) for linear chains in the blend with $N = 320$. By maintaining $L > 4R_{\mathrm{ee}}$, we preclude unphysical interactions between a chain and its own periodic images. Most importantly, by employing the generalized Kraynik-Reinelt boundary conditions[33–37] described below, this volume is sufficient to avoid such self-interactions even under uniaxial elongational flow. For the system studied here, we confirmed that the distance to self-periodic images was greater than $20\sigma$ even at the maximum Hencky strain of $\epsilon = 10$.

The systems were first equilibrated using Langevin dynamics at a constant temperature ($T = 1.0\frac{\varepsilon_{\mathrm{LJ}}}{k_{\mathrm{B}}}$). NEMD simulations were then performed to investigate the response under uniaxial elongational flow. The flow field was imposed by integrating the SLLOD equations of motion.[38] In standard elongational flow simulations, the simulation box rapidly contracts in the perpendicular directions, limiting the maximum achievable strain. To overcome this, we utilized the UEF package[35] and its extension for Langevin dynamics (UEFEX package),[37] which implements the generalized Kraynik–Reinelt boundary conditions[33,34] for long-time uniaxial deformation. The key strategy is the use of a triclinic simulation cell where the initial lattice vectors are tilted at a specific angle relative to the principal elongation axis.[33,34] By periodically re-mapping the deforming box back to its original tilted geometry—a technique conceptually similar to the original Kraynik–Reinelt boundary conditions[39–41]—the simulation can be continued indefinitely without the box dimensions ever falling below a critical threshold. This numerical "trick" enables us to maintain a consistent periodic boundary condition in all directions while imposing a constant strain rate up to very large Hencky strains ($\epsilon \geq 6$) [36,37], enabling the full observation of the thread-to-unthread transition in ring-linear blends.

The range of the imposed Hencky strain rates $\dot{\epsilon} \equiv \frac{\mathrm{d}\epsilon}{\mathrm{d}t}$ was selected to span the non-linear rheological regime, specifically within the interval $\frac{1}{\tau_{\mathrm{d}}} \lesssim \dot{\epsilon} \lesssim \frac{10}{\tau_{\mathrm{e}}}$, where $\tau_{\mathrm{d}}$ and $\tau_{\mathrm{e}}$ represent the longest relaxation time for chain reconfiguration (the disengagement time) and the relaxation time of segments between entanglement points (the entanglement time), respectively. In our model of flexible Kremer–Grest chains, the entanglement time is estimated to be $\tau_{\mathrm{e}} \approx 8000\tau$, which corresponds to $\tau_{\mathrm{d}}$ of $Z = 1$ ($N = 80$), while the values for the longest relaxation time $\tau_{\mathrm{d}}$ for each blend will be discussed and estimated in the following results section based on the orientational relaxation of the chain.[37] The selection of this range was also guided by numerical stability and precision requirements. For very high strain rates approaching or exceeding $\frac{10}{\tau_{\mathrm{e}}}$ (ten times the reciprocal of the entanglement time scale), we observed a significant rise in the system temperature due to intense artificial viscous heating, which complicates the maintenance of a stable isothermal state.[37] Furthermore, at lower strain rates near or below $\frac{1}{\tau_{\mathrm{d}}}$, the signal-to-noise ratio in the stress measurements becomes increasingly problematic, with large relative fluctuations in the transient stress making it difficult to resolve the fine features of the stress-strain curves.

Simulations were performed using HOOMD-blue[42] on GPUs for equilibrium analysis and LAMMPS[43,44] for non-equilibrium analysis. Molecular snapshots and visualizations were generated using OVITO.[45] To improve the signal-to-noise ratio in the transient stress measurements, the time-evolution data of the stress were smoothed using a Savitzky–Golay filter.[46]

# 3. Results

## 3.1. Linear Viscoelasticity and Relaxation Time

To establish the baseline rheological properties, we first examined the linear viscoelasticity of the ring-linear blends. **Figure 1** shows the shear relaxation modulus $G(t)$ for $N = 80$, 160, and 320 ($Z = 1$, 2, and 4, respectively). The shear relaxation modulus was calculated from the equilibrium stress fluctuations using the Green–Kubo relation.[47] To ensure high statistical precision even in the terminal relaxation regime, we employed the multiple-tau correlator method proposed by Ramírez et al.,[48] which allows for the efficient computation of the stress autocorrelation function over a wide range of time scales. A striking contrast is observed between the pure ring melts and the blends. The pure ring melts exhibit the fastest relaxation because ring polymers in their pure state cannot form conventional entanglements. In the intermediate time regime, the stress relaxation of these pure rings follows a power-law behavior with a power of 1/2, which can be understood simply as a manifestation of the self-similarity of Rouse chains. This power-law decay persists until the whole ring relaxes, leading into the terminal relaxation regime. According to experimental studies on pure ring polymer melts by Doi et al.,[49,50] the behavior of the shear relaxation modulus for systems with up to approximately $Z \approx 5$ can be well approximated by the Rouse ring model.

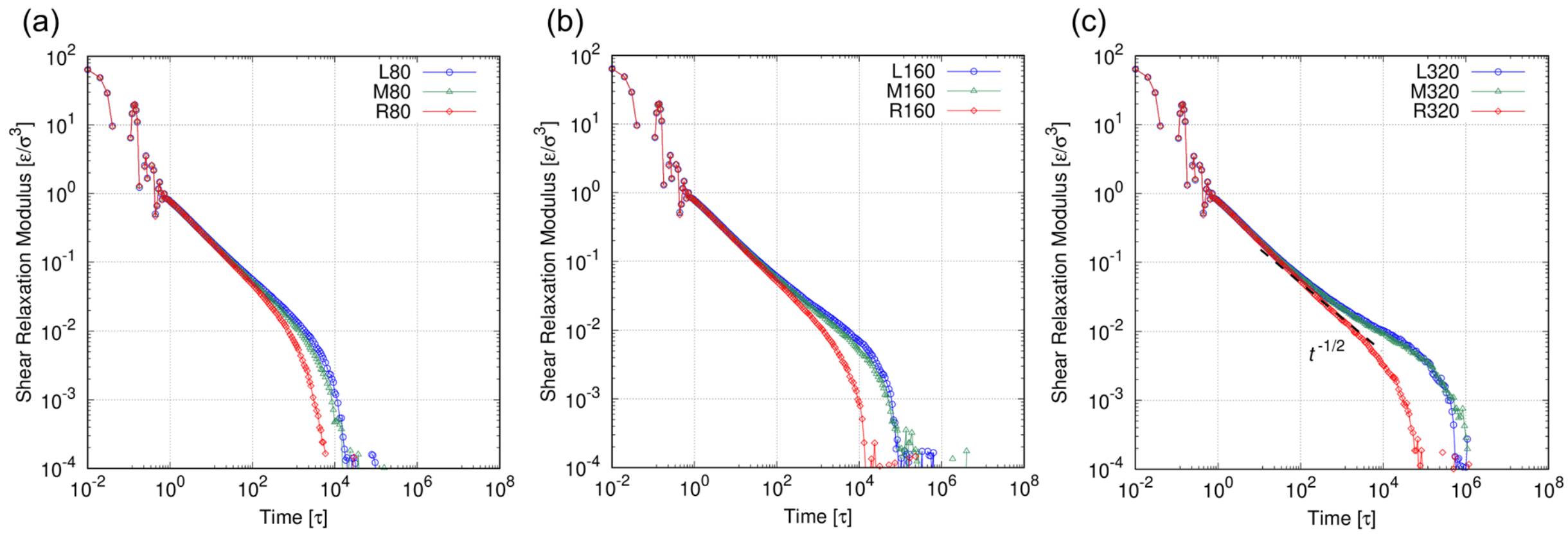

**Figure 1.** Shear relaxation modulus $G(t)$ as a function of time for (a) $N = 80$, (b) $N = 160$, and (c) $N = 320$. Each panel compares the linear melt (L), ring melt (R), and equi-weight ring-linear blend (M). The data were obtained from equilibrium coarse-grained molecular dynamics simulations using the Green–Kubo relation.

For the equi-weight blends, the relaxation behavior is remarkably similar to that of the pure linear melts. This behavior is consistent with previous experimental and simulation studies.[2,15] Specifically, for the systems with $N = 160$ and 320 ($Z = 2$ and 4), the stress relaxation evolves from an initial Rouse-like power-law (1/2) to a transient entanglement plateau before entering the terminal regime. This emergence of the plateau reflects the recovery of topological constraints: the threading of linear chains through the rings constraints the rings within the transient entanglement network formed by the linear matrix.[19] Consequently, the dynamics of the ring polymers are significantly retarded, becoming comparable to the relaxation of the linear chains as the threaded rings are effectively trapped within the matrix entanglements. For the $N =$ 320 blend, the relaxation of the 1:1 blend (M320) is nearly identical to that of the pure linear melt (L320), with the blend relaxing only slightly more slowly. This contrasts with the pronounced viscosity enhancement observed experimentally for symmetric ring-linear blends.[13] The difference can be understood from two features of the present system. First, our chains are only marginally entangled ($Z \approx 4$), unlike the well-entangled experimental systems. In this marginally entangled regime, threading—the microscopic origin of the viscosity enhancement in symmetric blends—is weak, since threading becomes ineffective for small rings.[13] Second, we employ the flexible Kremer–Grest model without bending stiffness, whereas chain stiffness is known to influence the threading behavior and hence the rheology of ring-linear blends. A direct quantitative comparison with experiment therefore requires longer and stiffer chains.

The characteristic relaxation times $\tau_d$ were determined by calculating the time correlation functions of characteristic vectors specific to the linear and ring architectures as follows. For linear chains, the correlation function of the end-to-end vector was used. For ring polymers, we calculated the correlation function of the diameter vector, defined as the vector connecting beads 1 and $N/2$ along the ring. The resulting correlation functions were fitted with the Kohlrausch-Williams-Watts (KWW) function, $\exp\{-(t/\alpha)^{\beta}\}$, to estimate the characteristic relaxation times, $\tau_d$, following the procedure described in our previous work.[37] These estimated relaxation times are summarized in **Table 1**. Note that $\tau_d$, $\alpha$, and $\beta$ are mathematically related via $\tau_d = \alpha/\beta\ \Gamma(1/\beta)$. Due to the strong nonlinearity of the gamma function, $\tau_d$ calculated from this relation is highly sensitive to small variations in $\beta$. When these parameters are rounded to three significant digits, the introduced rounding errors propagate significantly, causing a discrepancy between the reconstructed $\tau_d$ and the original values. Therefore, to ensure consistency between the tabulated parameters and the analytical relation, $\tau_d$ is presented with two significant digits, which accommodates the inherent numerical uncertainty of the fitting.

As shown in **Figure 2**, the relaxation times of each component exhibit distinct dependencies on chain length. The relaxation time of the linear component in the blend ($\tau_{d,M,L}$) remains nearly identical to that of the pure linear melt ($\tau_{d,L}$), suggesting that the transient entanglement network of the linear matrix is not significantly altered by the presence of rings at this composition. In contrast, while the relaxation time of the ring component in the blend ($\tau_{d,M,R}$) is relatively close to that of the pure ring melt ($\tau_{d,R}$) at small $N$, it deviates significantly as $N$ increases. For the longest chain length ($N$ = 320, $Z \approx 4$), $\tau_{d,M,R}$ shows a pronounced increase, becoming several times longer than $\tau_{d,R}$ and approaching the value of $\tau_{d,L}$. This trend indicates that as the chain length increases, the rings become more effectively trapped by the multiple threading of linear chains, causing their dynamics to be increasingly constrained by the surrounding transient entanglement network of the linear matrix. Comparing the data for $N$ = 160 and $N$ = 320 in **Table 1**, $\tau_{d,R}$ increases by a factor of 4, whereas $\tau_{d,M,R}$ increases by nearly a factor of 10. This signifies that the relaxation time of pure rings follows a scaling of $N^2$, while that of the rings in the ring-linear blend scales as $N^3$. These results are qualitatively consistent with previous studies on the dynamics of rings in linear matrices: the pronounced slowing down of ring dynamics due to threading by the surrounding linear chains has been reported experimentally using pulsed field gradient NMR spectroscopy by Kruteva et al.[11], in all-atom molecular dynamics simulations by Tsalikis et al.[7], and in coarse-grained simulations of tracer rings in linear matrices by Mo et al.[10]

Focusing on the KWW stretching exponent $\beta$ in **Table 1** reveals an interesting asymmetry. In the pure melts, the rings exhibit larger $\beta$ values than the linear chains of the same length ($N$ = 80: 0.618 vs. 0.602; $N$ = 160: 0.576 vs. 0.546; $N$ = 320: 0.520 vs. 0.459), indicating that they have a narrower relaxation-time spectrum than the linear chains. In the blends, however, the rings show smaller $\beta$ values than the corresponding pure-ring melts ($N$ = 80: 0.618 to 0.601; $N$ = 160: 0.576 to 0.525; $N$ = 320: 0.520 to 0.441), reflecting a broadening of the relaxation-time distribution. This suggests that threading by the linear chains diversifies the relaxation modes of the rings, increasing the complexity of their relaxation. The linear chains, by contrast, behave almost identically in the pure melts and in the blends, with nearly the same $\beta$ values for each chain length ($N$ = 80: 0.602 vs. 0.603; $N$ = 160: 0.546 vs. 0.541; $N$ = 320: 0.459 vs. 0.461). Their $\tau_d$ also differ by only a few percent, indicating that the relaxation behavior of the linear chains is essentially unaffected by the presence of rings. The effect of blending thus appears asymmetrically on the ring component, while the linear chains remain nearly undisturbed by the rings.

**Table 1.** Relaxation times for linear chains and rings in pure melts ($\tau_{d,L}$ and $\tau_{d,R}$), and blends ($\tau_{d,M,L}$ and $\tau_{d,M,R}$). The Kohlrausch-Williams-Watts functions' parameters, $\alpha$ and $\beta$, are also summarized. The entanglement time $\tau_e \approx 8000\tau$ was estimated from the value of $\tau_{d,L80}$.

| | $\tau_d$ [$\tau$] | $\alpha$ [$\tau$] | $\beta$ [−] |
|---|---:|---:|---:|
| L80 | 8000 | 5330 | 0.602 |
| R80 | 1800 | 1210 | 0.618 |
| M80,L | 7700 | 5120 | 0.603 |
| M80,R | 2400 | 1590 | 0.601 |
| L160 | 48000 | 27900 | 0.546 |
| R160 | 6700 | 4220 | 0.576 |
| M160,L | 50000 | 28500 | 0.541 |
| M160,R | 13000 | 7070 | 0.525 |

| L320 | 510000 | 213000 | 0.459 |
|---|---|---|---|
| R320 | 27000 | 14300 | 0.520 |
| M320,L | 500000 | 212000 | 0.461 |
| M320,R | 120000 | 45000 | 0.441 |

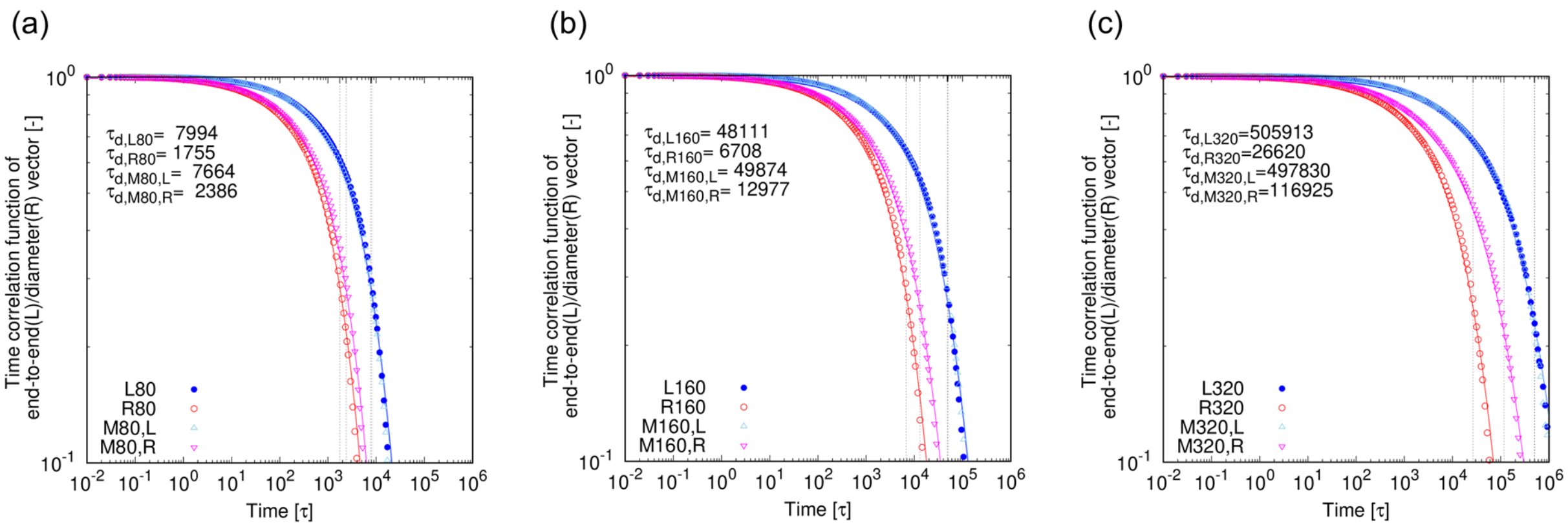

**Figure 2.** Time correlation functions of the end-to-end vector for linear chains (L) and the diameter vector for ring polymers (R) at (a) $N$ = 80, (b) $N$ = 160, and (c) $N$ = 320. The characteristic relaxation times ($\tau_d$) are indicated for each component in the pure melts and blends. The components in the blends are designated as M$N$,L and M$N$,R, corresponding to the linear and ring chains of length $N$.

## 3.2. Elongational Rheology

The non-linear rheological response of the ring-linear blends was investigated under uniaxial elongational flow. **Figure 3** shows the transient elongational viscosity $\eta_E(t) = N_1(t)/\dot{\epsilon}$ as a function of time for various strain rates $\dot{\epsilon}$, where $N_1(t) = \sigma_{zz}(t) - 0.5[\sigma_{xx}(t) + \sigma_{yy}(t)]$ is the first normal stress difference. A key finding of this study is the emergence of a distinct threshold in chain length for the appearance of stress overshoot in these blends. To evaluate the degree of non-linearity, we also plot the linear viscoelastic envelope $3\eta(t)$. This baseline is calculated from the shear relaxation modulus $G(t)$ obtained at equilibrium (shown in **Figure 1**) as: $3\eta(t) = 3\int_0^t G(s)ds$. In the limit of small strains or low strain rates, $\eta_E(t)$ is expected to follow the $3\eta(t)$ curve, consistent with Trouton's ratio.[51]

As observed in **Table 1**, the relaxation times of linear chains are nearly equivalent between the pure melt and the blend ($\tau_{d,L} \approx \tau_{d,M,L}$), while those of ring chains exhibit a substantial difference ($\tau_{d,R} < \tau_{d,M,R}$). Consequently, it is more straightforward to understand the characteristic time of the system by using the relaxation time of linear chains $\tau_{d,L}$ as a reference time of the Weissenberg number ($Wi = \dot{\epsilon}\, \tau_{d,L}$). Due to the large noise at lower strain rates, **Figure 3** presents the results for Weissenberg numbers of approximately 0.5 and above.

**Figure 3** displays the data smoothed via the Savitzky–Golay filter[46]. For comparison, the raw data alongside the smoothed results are presented in **Figure S1** in the Supporting Information. This supplementary figure also evaluates the degree of noise reduction when the system volume is increased eightfold (by doubling the length of each $x$, $y$, and $z$ dimension). Additionally, while **Figure 3** plots the results of a single run, the outcomes of multiple runs initiated from distinct initial configurations are provided in **Figure S2** in the Supporting Information. These multiple runs indicate that the initial-state dependence of the nonlinear behavior manifests primarily in the height of the overshoot peak, while the position of the peak and the subsequent steady-state value remain essentially unaffected.

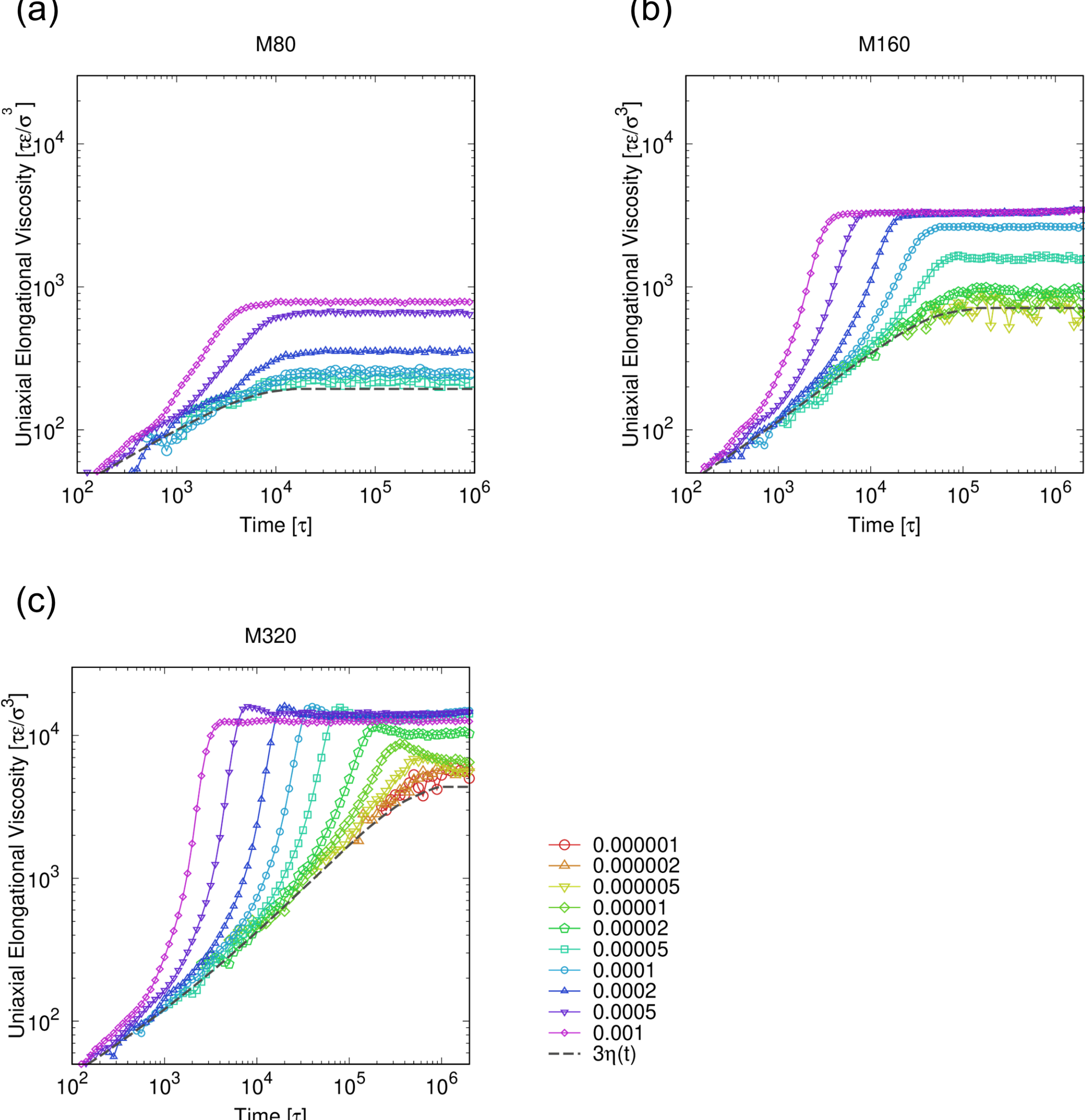

**Figure 3.** Transient uniaxial elongational viscosity $\eta_E(t)$ at various strain rates $\dot{\epsilon}$ for (a) M80 ($N$ = 80, $Z$ = 1), (b) M160 ($N$ = 160, $Z$ = 2), and (c) M320 ($N$ = 320, $Z$ = 4). A common legend is used for all three panels, where the symbols and their corresponding numerical values denote the applied strain rates, $\dot{\epsilon}$ [1/$\tau$]. The dashed line represents $3\eta(t)$, which serves as the linear viscoelastic envelope. Note that the strain rates investigated in this study correspond to Weissenberg number ranges of Wi = 0.8 to 8.0 ($\dot{\epsilon}$ = 0.0001 to 0.001), 0.48 to 48.0 ($\dot{\epsilon}$ = 0.00001

to 0.001), and 0.51 to 510.0 ($\dot{\epsilon}$ = 0.000001 to 0.001) for $N$ = 80, 160, and 320, respectively. The distinct stress overshoots are observed only for the M320 case when $\dot{\epsilon} > 1/\tau_d$ (Wi > 1), identifying the threshold chain length for non-linear extensional rheology.

As the strain rate increases and exceeds the reciprocal of the longest relaxation time (Wi > 1), all systems started to exhibit significant strain hardening, where the elongational viscosity grows rapidly beyond the linear $3\eta(t)$ envelope. However, the subsequent qualitative behavior depends strongly on the chain length. For the blends with shorter chains, M80 ($N$ = 80, $Z$ = 1) and M160 ($N$ = 160, $Z$ = 2), the elongational viscosity exhibits monotonic growth followed by a steady state, as shown in **Figures 3a and b**. Even at high strain rates where the chains are expected to be significantly deformed and stretched, no macroscopic stress overshoot was observed in the total stress of these systems. In sharp contrast, the M320 blend ($N$ = 320, $Z$ = 4) displays a clear and pronounced stress overshoot following the initial strain hardening, as shown in **Figure 3c**. This transition from monotonic growth to overshoot behavior as the chain length increases from $Z$ = 2 to 4 identifies $Z \approx 4$ as the threshold degree of threading required to trigger non-linear stress signatures in ring-linear blends. This threshold suggests that a minimum number of topological constraints—specifically, multiple threading of linear chains through a single ring—is necessary to sustain a sufficiently high conformational tension that can be rapidly released upon unthreading, leading to the observed stress peak. Following this logic, one would expect the stress overshoot to be even more pronounced in systems with higher degrees of entanglement. To test this, we further examined the case of $Z$ = 8 ($N$ = 640) and indeed observed a clear stress overshoot (**Figure S3** in the Supporting Information), as expected. These results collectively demonstrate that the stress overshoot is not a phenomenon exclusive to specific chain architectures or rigidities, but can be clearly manifested even in flexible chain systems once the degree of entanglement exceeds a certain threshold.

To further clarify the molecular origin of the stress overshoot observed for the M320 blend ($N$ = 320, $Z$ = 4), we decomposed the total first normal stress difference $N_1$ into the contributions of the ring and linear components. **Figure 4** shows this decomposition as a function of Hencky strain $\epsilon = t\dot{\epsilon}$ at various strain rates $\dot{\epsilon}$. The stress components for the ring and linear chains were determined by accumulating the per-atom stress values calculated for each respective chain type. As shown in **Figure 4a**, the total stress exhibits a peak before reaching a steady state, particularly at higher strain rates. The decomposition reveals that this overshoot behavior is primarily driven by the ring component, as shown in **Figure 4b**. The stress contribution from the rings increases sharply, reaches a distinct maximum at a Hencky strain of 3–4, and then significantly relaxes. Notably, this peak is more clearly observable at intermediate strain rates rather than the highest one. This occurs because, at lower strain rates, the flow-induced constraints are less dominant, allowing the rings to relax more effectively. Consequently, the stress peak stands out more prominently than at the highest rate. In contrast, as shown in **Figure 4c**, the linear component exhibits a monotonic increase in stress followed by a plateau, with no detectable overshoot.

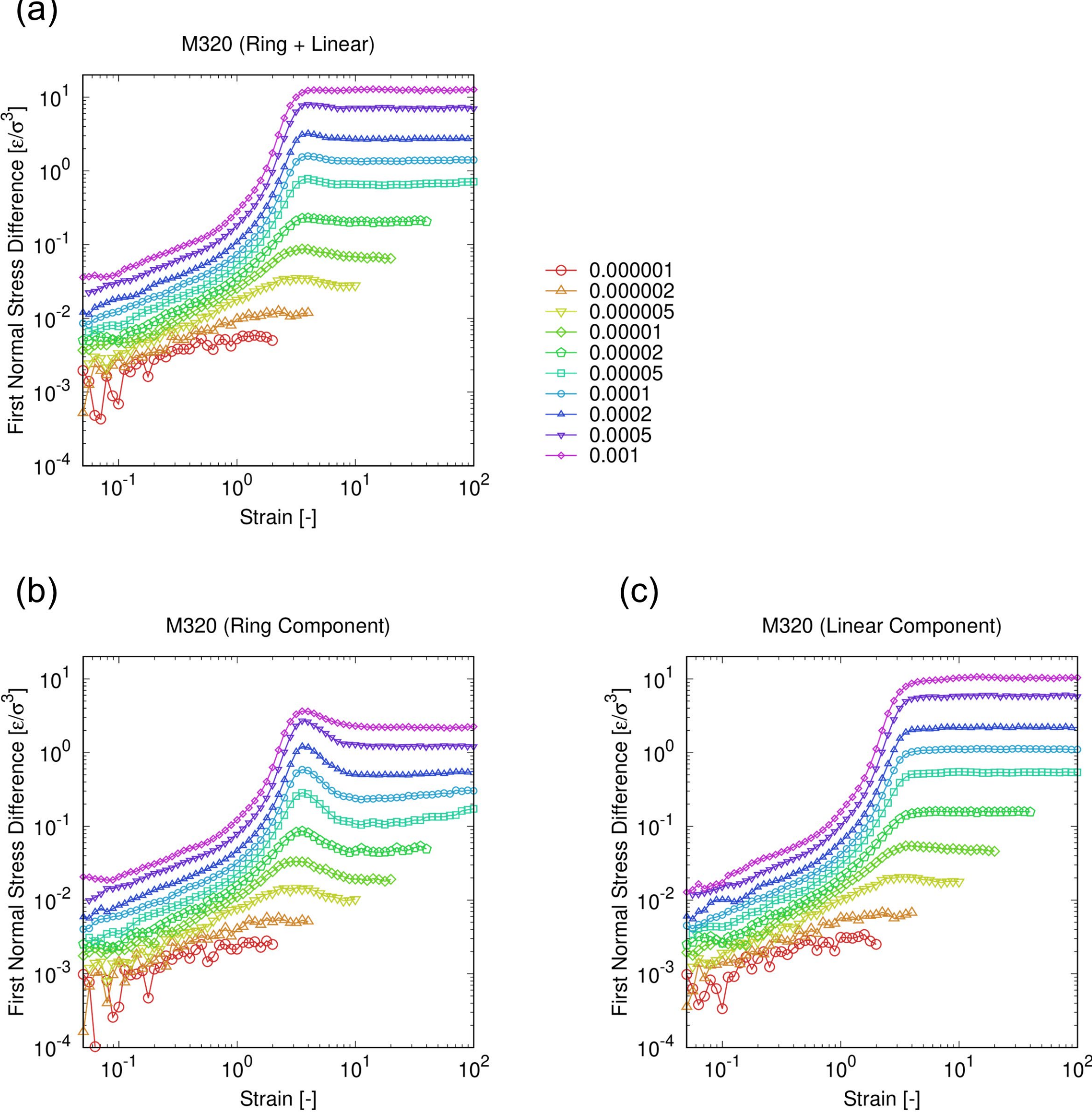

**Figure 4.** Decomposition of the first normal stress difference $N_1 = \sigma_{zz} - 0.5(\sigma_{xx} + \sigma_{yy})$ for the M320 blend ($N$ = 320, $Z$ = 4) as a function of Hencky strain. (a) Total stress of the ring-linear blend, (b) contribution from the ring component, and (c) contribution from the linear component. The results show that the stress overshoot is primarily driven by the ring component.

This emergence of a stress peak in the ring component is highly dependent on the chain length. The corresponding data for $N = 80$ and 160 are provided in the Supporting Information (**Figures S4 and S5**). As shown in **Figure S4b** for the M160 blend ($Z = 2$), the ring component exhibits only a subtle stress overshoot at the higher strain rates. However, its magnitude is insufficient to induce a clear overshoot in the total stress of the blend (**Figure S4a**), where monotonic growth remains dominant. For the even shorter M80 blend ($Z = 1$), no overshoot is observed in either the ring or the linear component, as shown in **Figure S5**, with both exhibiting purely monotonic stress growth. The comparison between **Figures 4**, **S4**, and **S5** highlights that while rings have an inherent tendency to exhibit overshoot when blended with linear chains, a threshold chain length of approximately $Z \approx 4$ is required for the threading to be sufficiently persistent to produce a macroscopic rheological signature.

Following the stress component decomposition presented in **Figure 4**, **Figure 5** provides a more detailed analysis by further decomposing the stress into specific inter-molecular contributions: ring–ring, linear–linear, and ring–linear interactions. The ring–linear cross term represents the quantity that was equally partitioned (1/2 each) into the individual ring and linear components presented in **Figure 4**. While those ring and linear components (**Figure 4**) could be computed during the simulation run simply by identifying the particle species, calculating the specific ring–ring, ring–linear, and linear–linear interactions requires distinguishing the exact pairs of interacting particles (and further decomposing them into forces acting between different molecules). The stress was evaluated from the per-atom virial, and each contribution was assigned according to the species of the interacting particles: bonded interactions, being intramolecular, contribute only to the same-species (ring–ring or linear–linear) terms, whereas nonbonded interactions are assigned to the ring–ring, ring–linear, or linear–linear term according to the molecular species of the two partners. To avoid the computational time loss associated with this pairs-identification during the simulation, it is more straightforward to analyze these terms from saved snapshots. We first confirmed that the snapshot-derived data successfully reproduced the original continuous behavior of the total stress as shown in **Figure 5**. Having validated this approach, we obtained the stress profiles for the ring–ring, ring–linear, and linear–linear components. Both the ring–linear and ring–ring components exhibit a distinct stress overshoot, suggesting that the threading between ring and linear chains, as well as between ring and ring chains, triggers this overshoot behavior. Interestingly, while the ring–linear component retains a finite stress value after the overshoot, the ring–ring component decays to nearly zero. This contrast indicates that the rings become surrounded by the linear chains after the overshoot.

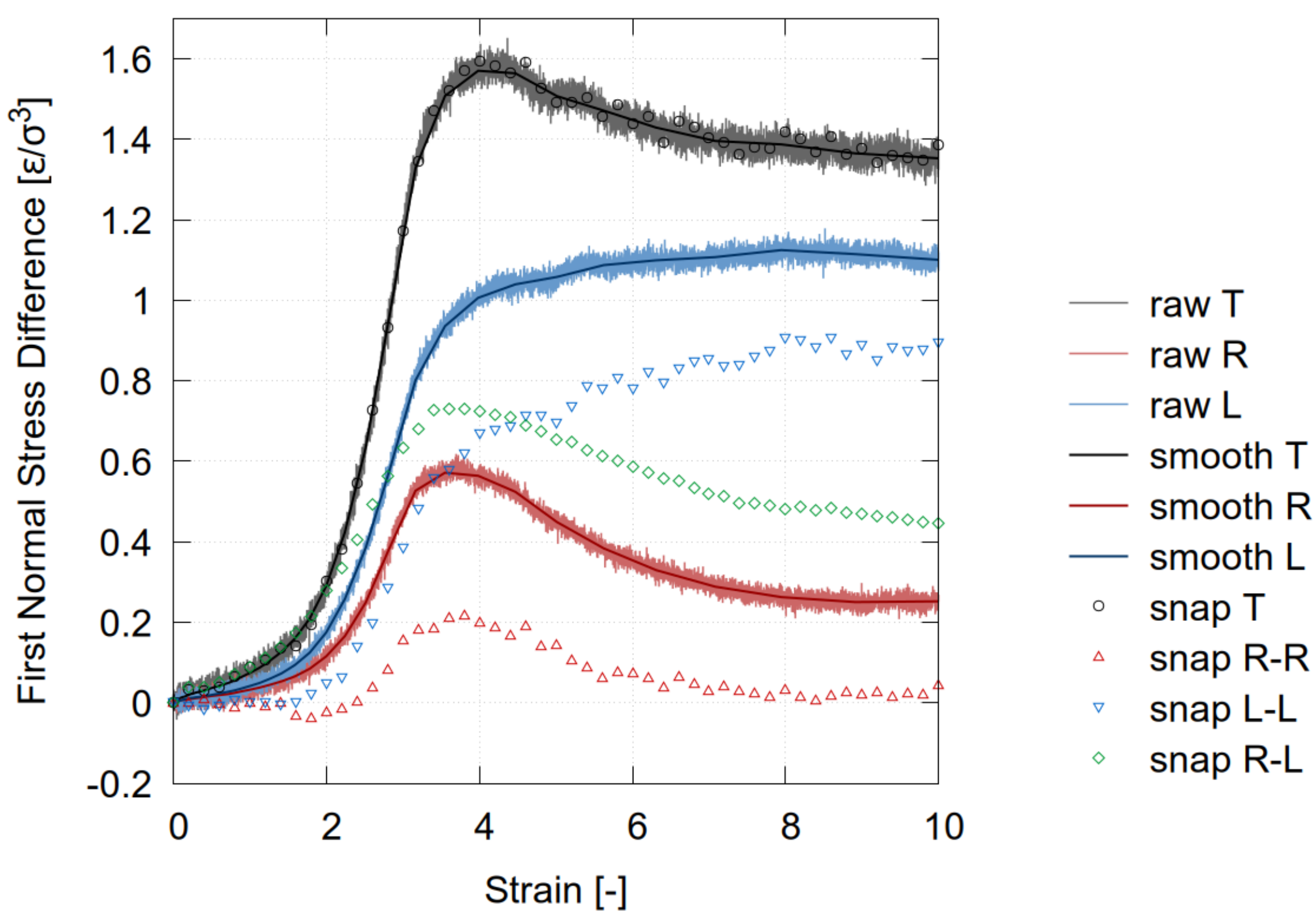

**Figure 5.** The first normal stress difference for the M320 of ring-linear blend under a strain rate of 0.0001 as a function of Hencky strain. Solid lines represent the continuous raw and smoothed profiles for the total (T), ring (R), and linear (L) components. Open circles (snap T) show the total stress recalculated from saved snapshots to validate the accuracy of the post-processing. The symbols (triangles and diamonds) present the further breakdown of the stress into specific pair interactions (ring–ring (snap R–R), linear–linear (snap L–L), and ring–linear (snap R–L)).

To bridge the gap between macroscopic stress behavior and microscopic chain dynamics, we analyzed the conformational changes of the polymer chains during elongational flow. In the following analysis, we focus on the results obtained at a strain rate of $\dot{\epsilon}$ = 0.0001, as this rate corresponds to the regime where the stress overshoot is prominently observed, allowing for a clear characterization of the structural transitions. **Figure 6** shows snapshots and the probability distributions of the span length (maximum intramolecular distance) for the M320 blend at various Hencky strains. Here, the span length is defined as the largest separation between any two beads within a single chain, serving as a robust indicator of chain extension; a value close to the contour length signifies a fully stretched, rod-like conformation.

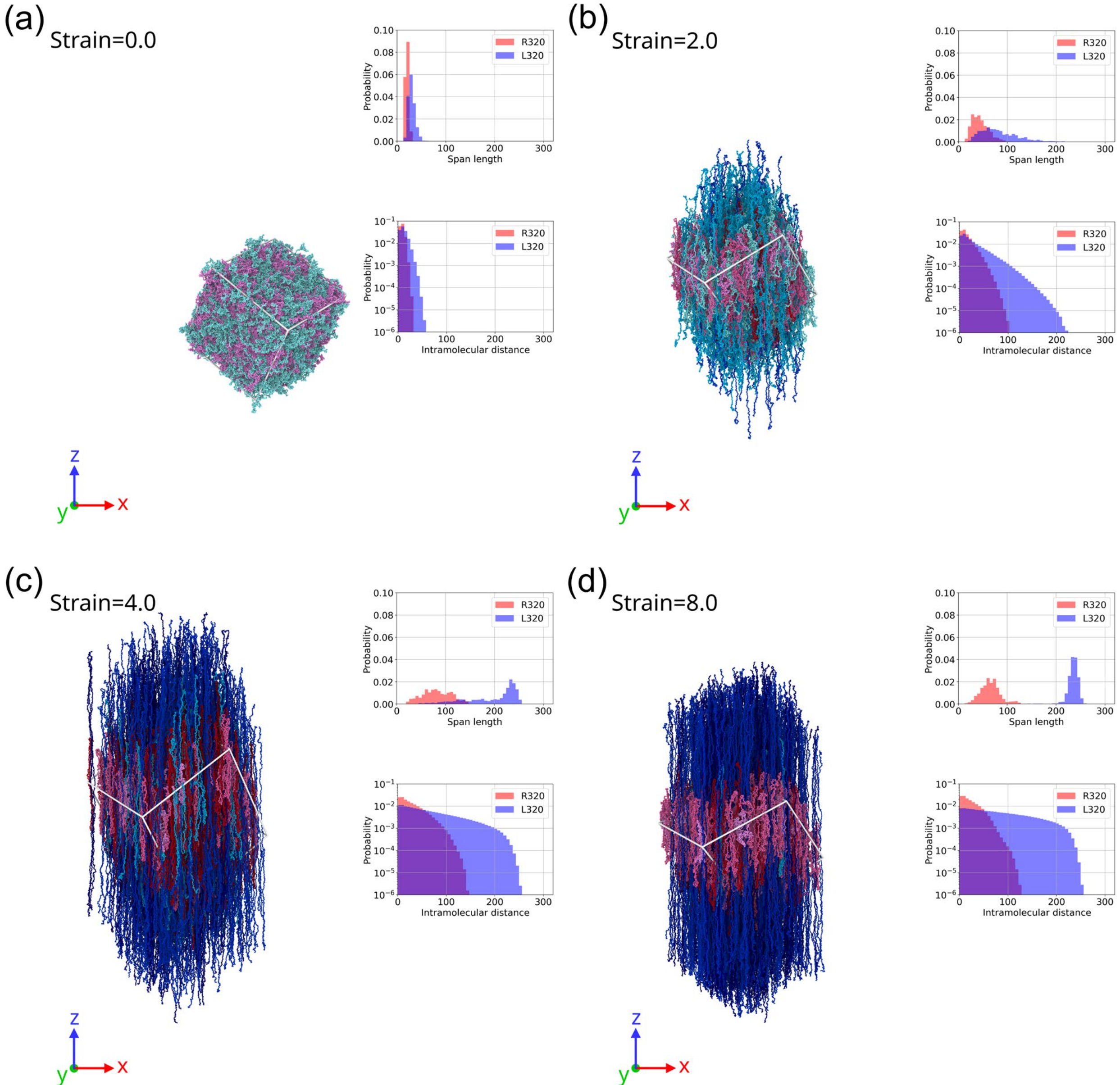

**Figure 6.** Chain conformation analysis and snapshots of the M320 ring-linear blend under $\dot{\epsilon}$ = 0.0001 at Hencky strains of (a) 0.0, (b) 2.0, (c) 4.0, and (d) 8.0. The white frames correspond to the unit cell shape at each strain. Each chain is unwrapped to maintain its structural continuity and is positioned such that its center of mass resides within the unit cell. In the snapshots, ring and linear polymers are represented in reddish and bluish color scales, respectively. The color intensity of each individual chain was assigned based on its span length: lighter shades

correspond to smaller distances, while darker shades indicate larger distances. The accompanying plots show the probability distribution of the span length and the individual intramolecular distance for ring (R320) and linear (L320) chains.

In the snapshots, ring and linear polymers are colored in reddish and bluish scales, respectively, with darker shades indicating more highly stretched chains (larger span length). At equilibrium (**Figure 6a**, strain of 0.0), both components exhibit coiled conformations with a narrow distribution of intramolecular distances. As the strain increases to 2.0 and 4.0 (**Figures 6b and c**, respectively), the distributions shift significantly toward larger distances, and the snapshots show a clear increase in dark-shaded chains, indicating that both rings and linear chains are being stretched along the elongational direction. Notably, as the stress increases further (strain ≈ 4), the ring component shows a wide distribution extending to the maximum possible distance, confirming that many rings are in a highly stretched state due to the strong topological constraints imposed by the multiple threading of linear chains. However, at a higher strain of 8.0 (**Figure 6d**), while the linear chains remain in a highly stretched steady state, the distribution for the ring component slightly shifts back toward smaller distances, and the intensity of the dark-red chains decreases. This visual evidence supports the thread-to-unthread transition: after reaching a maximum stretch, the ring polymers contract as the linear chains eventually unthread, resulting in the relaxation of the ring's conformational tension and the subsequent stress overshoot observed in **Figure 4**.

# 4. Discussion

## 4.1. Topological Constraint Analysis via Gauss Linking Number

To provide quantitative evidence for the thread-to-unthread transition, we performed a statistical analysis of the topological constraints using the Gauss linking number. Because the Gauss linking number is strictly defined for a pair of closed loops, we applied a numerical procedure to evaluate the threading between a ring and a linear chain. As an inexpensive pre-screening, we first defined a bounding sphere for each ring and each linear chain from its center of mass and the maximum distance of its beads from that center, and retained only those ring-linear pairs whose bounding spheres overlap (under the minimum-image convention for the triclinic box) as candidates for threading; pairs whose bounding spheres do not overlap cannot be threaded and were discarded. For each candidate pair, we artificially closed the linear chain into a loop by extending its two ends in the direction away from the center of mass and connecting them in a large arc.[8] This arc was constructed outside the entire ring-linear pair, so that it always encloses the whole pair and can never penetrate the ring; the resulting Gauss linking number is therefore well-defined and independent of the closure geometry. In this quasi-ring and ring system, if the original linear chain had threaded through the ring, the resulting pair formed a catenane-like structure with a Gauss linking number of 1. The Gauss linking number was evaluated using the Topoly library.[52]

**Figure 7** shows the probability distribution of the number of penetrations per ring obtained using this method. While the results discussed thus far have been based on a baseline strain rate of $\dot{\epsilon} = 0.0001$, we examined the $N = 80$ (M80) blend at a higher rate of $\dot{\epsilon} = 0.001$ because the former rate was below the reciprocal of its longest relaxation time ($1/\tau_d$), placing it in the linear viscoelastic regime. To enable a systematic comparison across different chain lengths at identical flow strengths, the results for the $N = 160$ (M160) blend are presented at both strain rates. This allows for a direct comparison between $N$ = 320 and 160 at $\dot{\epsilon}$ = 0.0001 (**Figures 7a and b**), and between $N$ = 160 and 80 at $\dot{\epsilon}$ = 0.001 (**Figures 7c and d**).

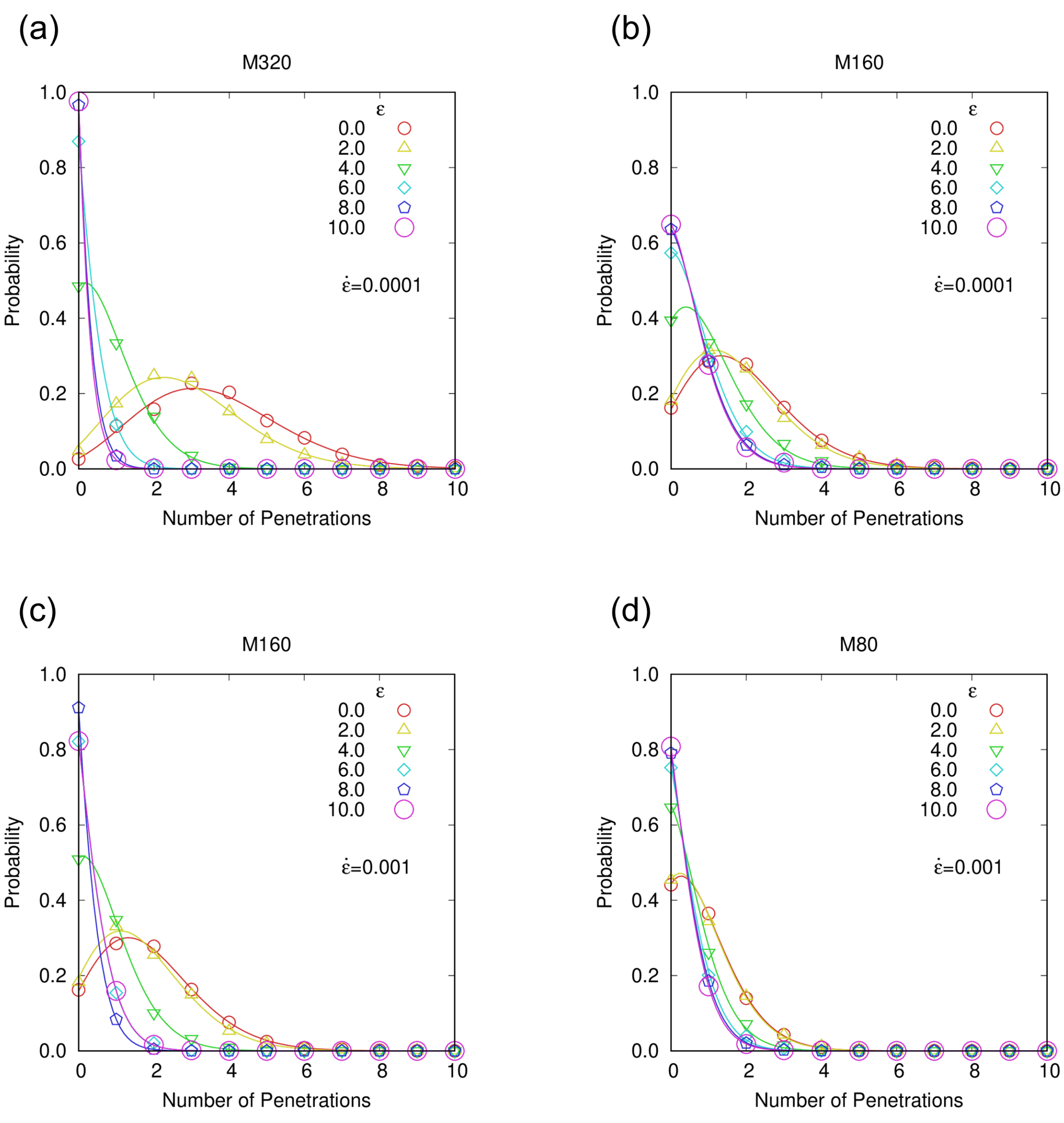

**Figure 7.** Probability distribution of the number of penetrations (threading) per ring for (a) M320 and (b) M160 at a low strain rate ($\dot{\epsilon}$ = 0.0001) and (c) M160 and (d) M80 at a high strain rate ($\dot{\epsilon}$=

0.001). Threading was evaluated using the Gauss linking number. The results indicate that the number of penetrations decreases as strain increases, triggering the unthreading transition.

For the M320 blend (**Figure 7a**), each ring is penetrated by multiple linear chains at equilibrium (strain of 0.0), with the distribution peaking at approximately three to four penetrations. This confirms the rings are well constrained within the transient entanglement network of the linear matrix through threading. As the Hencky strain increases, the distribution shifts significantly toward zero. At a strain of 4.0, which corresponds to the stress peak, the peak of the distribution has already moved to a lower number of penetrations. By a strain of 8.0, the majority of rings have zero penetrations, indicating that most linear chains have unthreaded from the rings. This dramatic reduction in topological constraints directly correlates with the relaxation of the ring component's stress observed in **Figure 4b**. In contrast, for shorter chains such as M160 (**Figures 7b** and **c**) and M80 (**Figure 7d**), the average number of penetrations at equilibrium is much lower. For M80, most rings have either zero or only one penetration even at equilibrium. The comparison across **Figures 7a–d** suggests that a high initial degree of multiple threading is a necessary condition for sustaining sufficient conformational tension to produce a macroscopic stress overshoot. The rapid release of these multiple constraints during elongational flow is the fundamental molecular mechanism driving the observed non-linear rheological signature in the M320 blend.

## 4.2. Direct Visualization of the Thread-to-Unthread Transition

To provide a more intuitive and microscopic view of the transition mechanism, we monitored the structural evolution of a representative ring-linear complex. **Figure 8** displays a series of snapshots for a specific configuration of the M320 blend, where a single ring (colored in red) is initially threaded by four linear chains (shown in lighter colors). To clarify the dynamics of this specific cluster, the chains were shifted such that their centers of mass remain within the periodic simulation cell throughout the deformation process. At the onset of flow and up to a Hencky strain of 4.0 (**Figures 8a–c**), the four penetrating linear chains act as temporary topological cross-links. As the linear chains are stretched by the elongational flow, they drag the ring along the flow direction, forcing it into a highly extended, near-rod-like conformation. Specifically, at the strain of 4.0, the ring is threaded by two linear chains that exert frictional forces in opposite directions (along the positive and negative $z$-axes), effectively stretching the ring open. This state corresponds to the peak of the ring's stress contribution observed in **Figure 4b**. However, as the strain further increases to 8.0 (**Figure 8d**), the linear chains are observed to eventually unthread or slip out from the ring's cavity. Once the topological constraints, acting as an internal expansive force on the ring, are released, the ring is no longer supported by the tension of the linear chains and spontaneously contracts to a more relaxed, mildly stretched state. This direct visualization of the thread-to-unthread transition confirms that the macroscopic stress overshoot is not merely a collective effect but is rooted in the dynamic release of multiple threading constraints at the individual chain level.

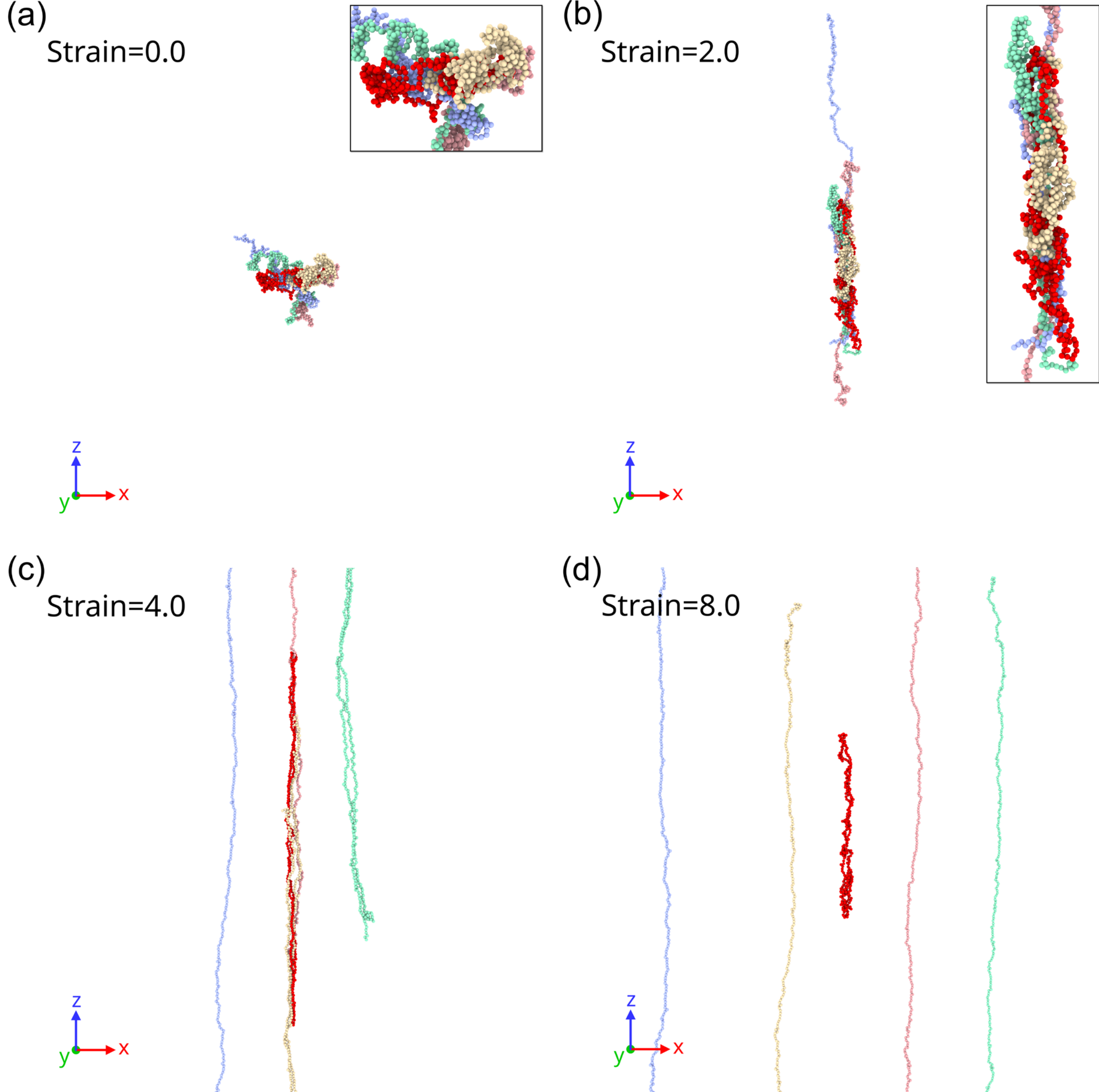

**Figure 8.** Representative visualization of the thread-to-unthread transition under uniaxial elongational flow for the M320 blend at $\dot{\epsilon} = 0.0001$. For clarity, a specific configuration where four linear chains penetrate a single ring was selected for visualization. Snapshots at Hencky strains of (a) 0.0, (b) 2.0, (c) 4.0, and (d) 8.0 illustrate the dynamic process: the ring (red) is significantly stretched by the multiple threading of linear chains (light colors) before contracting as the linear chains unthread. To maintain the integrity of the molecular structures, the positions

of the chains were shifted such that their centers of mass remain within the periodic simulation cell. For (a) and (b), magnified views of the ring–linear threading region are shown as insets.

### 4.3. Real-Space Probability Density Analysis

To further corroborate the ring contraction following the stress overshoot, we evaluated the real-space probability density of the segments for the ring and linear components. To compute these profiles, we first mapped the segment positions onto a three-dimensional grid with a mesh size of $1\sigma$ using linear interpolation. The density values were calculated by first aligning the center of mass of each chain at the origin and then averaging over all chains of the same species and multiple independent configurations. The resulting probability density is normalized such that its integral over the entire volume equals the total number of segments $N$ per chain. **Figure 9** shows the density profiles along the elongational direction ($z$ axis) and the perpendicular directions ($x$ or $y$) at various Hencky strains for the M320 blend. Because the system maintains symmetry in the $x$–$y$ plane perpendicular to the elongational axis, the density profiles in these directions were calculated by taking a cylindrical average around the $z$ axis to improve statistical accuracy. To provide the further structural details, **Table 2** summarizes the radius of gyration ($R_g$) and its components along $z$ axis ($R_{g,z}$) and perpendicular direction ($R_{g,\perp} = \sqrt{R_{g,x}^2 + R_{g,y}^2}$) for both rings and linear chains in the M320 blend.

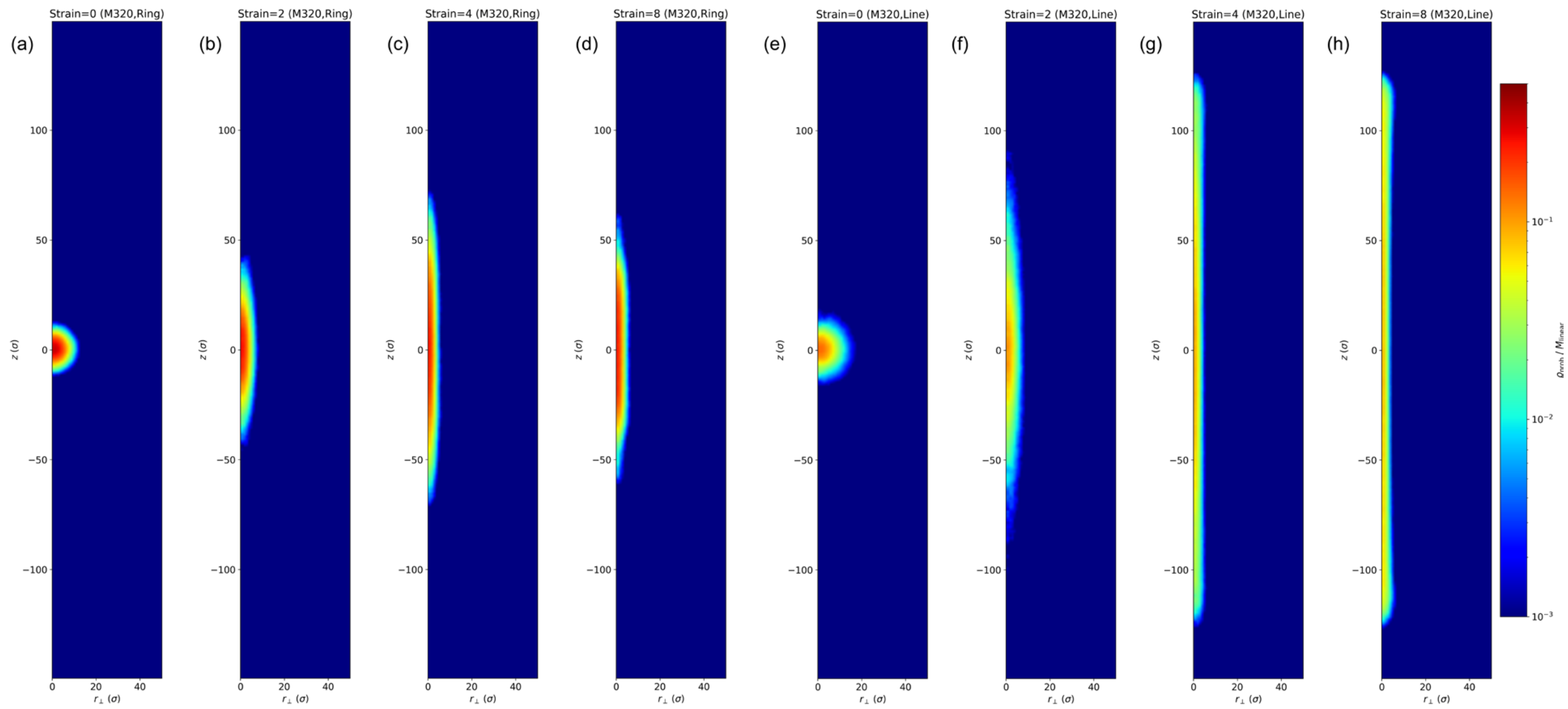

**Figure 9.** Real-space probability density of ring and linear components in the M320 blend ($\dot{\epsilon}$ = 0.0001) at Hencky strains of (a) 0.0, (b) 2.0, (c) 4.0, and (d) 8.0 for ring components, and (e) 0.0, (f) 2.0, (g) 4.0, and (h) 8.0 for linear components. Here, $r_{\perp} = \sqrt{x^2 + y^2}$.

**Table 2.** Radius of gyration ($R_g$) and its components along $z$ axis ($R_{g,z}$) and perpendicular direction ($R_{g,\perp} = \sqrt{R_{g,x}^2 + R_{g,y}^2}$) for rings and linear chains in the M320 blend. Note that $R_g^2 = R_{g,z}^2 + 2R_{g,\perp}^2$.

| | Topology | Strain = 0 | Strain = 2 | Strain = 4 | Strain = 8 |
|---|---|---|---|---|---|
| $R_g$ | Ring | 6.3$\sigma$ | 13.1$\sigma$ | 24.5$\sigma$ | 19.9$\sigma$ |
| | Linear | 9.4$\sigma$ | 25.7$\sigma$ | 59.8$\sigma$ | 71.6$\sigma$ |
| $R_{g,z}$ | Ring | 3.5$\sigma$ | 12.6$\sigma$ | 24.3$\sigma$ | 19.7$\sigma$ |
| | Linear | 5.1$\sigma$ | 25.3$\sigma$ | 59.7$\sigma$ | 71.6$\sigma$ |
| $R_{g,\perp}$ | Ring | 3.6$\sigma$ | 2.3$\sigma$ | 1.6$\sigma$ | 1.7$\sigma$ |
| | Linear | 5.3$\sigma$ | 2.8$\sigma$ | 1.7$\sigma$ | 1.6$\sigma$ |

Consistent with the conformational analysis of **Figure 6**, the ring component exhibits a dramatic expansion and subsequent contraction in real space. At the stress peak (strain of 4.0), the probability density of the ring segments shows its broadest distribution along the $z$ axis, indicating that the rings are stretched to their maximum spatial extent along the elongational direction. However, as the strain increases to 8.0, the distribution width of the ring component along the $z$ axis decreases, accompanied by an increase in the peak density at the center. This spatial contraction serves as a direct structural signature of the unthreading process, where the release of multiple topological constraints allows the ring to recoil despite the continuous elongational flow. In contrast, the linear component displays a different trend. As shown in the corresponding density profiles, the linear chains reach a maximum degree of extension by $\epsilon = 8.0$. This lack of contraction in the linear component reinforces our conclusion that the stress overshoot is a phenomenon unique to the ring polymers in the blend, driven by the dynamic transition from a threaded, highly stretched state to an unthreaded, relaxed state.

At a strain rate of $\dot{\epsilon} = 0.0001$, which is sufficiently fast compared to the reciprocal of the longest relaxation times of both rings and linear chains, both components reach a highly stretched steady state. In this steady state, the average extension of each chain is generally determined by the balance between the outward drag force from the elongational flow and the inward entropic tension of the chain itself. However, the presence of the stress overshoot in the ring component suggests that there is an additional mechanism at play. The microscopic origin of the overshoot lies in the "extra stretching" of the ring polymers. During the initial stages of deformation, the linear chains that penetrate the ring act as moving constraints that actively pull the ring along the flow direction. This coupling generates a transient extension that exceeds the degree of stretching by flow-drag alone. As the linear chains eventually unthread, this extra pulling force vanishes, and the ring contracts to its steady-state length where only the flow-drag and entropic tension are balanced. This structural transition is clearly captured in **Figures 9c and d**, where the $z$ axis distribution of the ring component at $\epsilon = 4.0$ (the stress peak) is broader than at $\epsilon = 8.0$ (the steady state).

## 4.4. Predicted SANS Patterns for Experimental Validation

Finally, we propose an experimental approach to validate the thread-to-unthread transition using SANS. As suggested by **Figure 9**, the average conformation of the rings is expected to be strain dependent. There is a strong motivation to capture this transition through direct experimental observation; indeed, SANS measurements on labeled rings[12,13,53–56] have already proven to be a viable method for this purpose. In a previous experimental study, Borger et al.[22] utilized SANS to investigate the same phenomenon. Specifically, they employed a labeling strategy in which 10% of deuterated linear chains were incorporated as tracers to monitor the conformational changes of the linear component. By comparing the scattering profiles of the ring-linear blend with those of a pure linear melt, they aimed to isolate the effects of the ring-linear topological interactions. This experimental design was strategically chosen to detect how the presence of rings induces extra stretching within the transient entanglement network of the linear matrix. Measuring the ring component directly under flow remains experimentally challenging. In particular, isolating the scattering contribution of the ring component by substituting 10% of the rings with their deuterated counterparts is a significant challenge. While some specialized studies have successfully employed deuterated rings for SANS measurements[12,13,53–56], the prohibitive cost and the synthetic difficulty of ensuring the absence of linear contaminants when scaling up production to the levels required for precise structural analysis under flow have largely restricted more widespread validation. Our simulation, free from such contrast constraints, provides a complementary perspective by predicting the structural evolution of the ring component itself. While Borger et al.[22] provided indirect evidence of the transition through the deformation of the linear matrix, our results offer a more direct structural signature: the post-peak relaxation of ring anisotropy.

The predicted scattering intensity $I(\boldsymbol{q})$ was calculated based on the total coherent scattering from the $N$ segments of each chain as $I(\boldsymbol{q}) = \left\langle \left| \sum_{j=1}^{N} \exp\left(i\boldsymbol{q} \cdot \boldsymbol{r}_j\right) \right|^2 \right\rangle$, where $\boldsymbol{q}$ is the wave vector and $\boldsymbol{r}_j$ is the position of the $j$-th segment. Here, the angular brackets $\langle \cdots \rangle$ denote the ensemble average over all chains of the same topology (i.e., either linear or ring). This definition ensures that the scattering intensity properly reflects the overall size and shape of the molecules, with $I(0) = N^2$ in the limit of small $q$. To improve statistical accuracy and account for the symmetry of the uniaxial elongational flow, the 2D SANS patterns in the $q_z$–$q_\perp$ plane were obtained by performing a cylindrical average in $q$-space ($q_\perp = \sqrt{q_x^2 + q_y^2}$), consistent with the real-space probability density analysis. **Figure 10** shows the predicted 2D SANS patterns for the M320 blend.

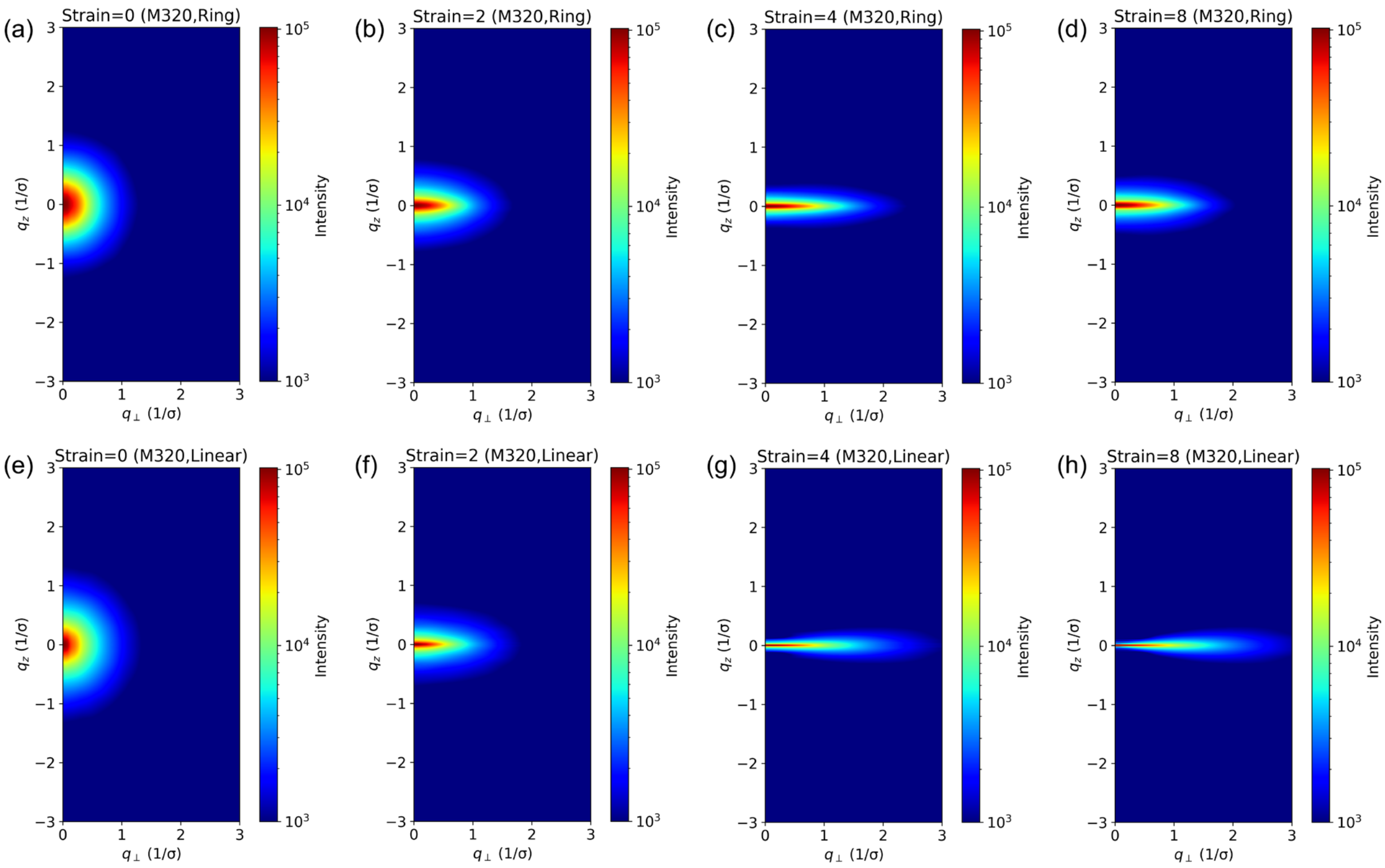

**Figure 10.** Predicted 2D SANS patterns for experimental validation of the unthreading mechanism. Patterns (a–d) correspond to the ring component and (e–h) to the linear component in the M320 blend ($\dot{\epsilon}$ = 0.0001) at various strains. The ring component shows maximum anisotropy at the stress peak ($\epsilon$ = 4.0) and a shift toward a more relaxed state at $\epsilon$ = 8.0 due to unthreading.

For the ring component (**Figures 10a–d**), the scattering pattern exhibits a dramatic evolution that corresponds to the stress overshoot. As the strain increases, the scattering intensity becomes increasingly localized near $q_z$ = 0 while simultaneously extending toward the high-$q_\perp$ region. Notably, at the stress peak ($\epsilon$ = 4.0), this anisotropy reaches its maximum, where the intensity is highly localized near $q_z$ = 0. This narrow width in $q_z$ reflects the maximally extended, rod-like conformation of the rings held by multiple threading constraints as shown in **Figure 8**. Beyond the stress peak, as the rings become unthreaded, the scattering intensity starts to spread along the $q_z$ direction, signaling the onset of structural relaxation. By $\epsilon$ = 8.0, the distribution in $q_z$ becomes notably broader than at $\epsilon$ = 4.0, which is consistent with the spatial contraction of the rings observed in **Figures 9c–d**. This shift from a high anisotropic state to a more relaxed state following the stress peak serves as a structural signature of the unthreading process.

In contrast, the linear component (**Figures 10e–h**) maintains a highly anisotropic pattern from $\epsilon$ = 4.0 to 8.0, reflecting its steady-state extension. Specifically, the scattering intensity is sharply concentrated at $q_z \approx 0$ in the low-$q_\perp$ region, indicating a significant increase in the correlation length along the $z$ axis as the chains extend and align with the stretching direction. Meanwhile, the pattern exhibits a broader distribution along the $q_z$ direction at higher $q_\perp$, which can be attributed to local segmental fluctuations that deviate from the perfectly oriented backbone.

To further investigate the structural anisotropy, 1D intensity profiles were extracted from the 2D-SANS patterns along the $q_z$ and $q_\perp$ axes shown in **Figure S6** in the Supporting Information. Guinier analysis of these 1D profiles (**Figure S7**) yielded the radius of gyration components, $R_{g,z}$ and $R_{g,\perp}$which are summarized in **Table S1**. These values were found to be in good agreement with the values calculated directly from the coordinate data presented in **Table 2**, confirming the validity of our analysis.

To extract fine anisotropic features of the predicted 2D SANS patterns, we applied the spherical harmonic expansion technique[57,58], which has been used to analyze 2D scattering patterns of entangled polymer melts under uniaxial elongational flow. The 2D patterns $I(\boldsymbol{q})$ were expanded into spherical harmonic components $S_l^m(q)$; $I(\boldsymbol{q}) = \sum_{l,m} S_l^m(q) Y_l^m(\hat{\boldsymbol{q}})$, where $Y_l^m(\hat{\boldsymbol{q}})$ is the spherical harmonic function and $\hat{\boldsymbol{q}} = \boldsymbol{q}/q$ specifies the direction of the wave vector $\boldsymbol{q}$ on the unit sphere. Because of the uniaxial symmetry of the flow, we take the elongation (stretching) axis as the polar ($z$) axis of the spherical harmonic expansion. The flow then preserves rotational symmetry about this axis, so the leading anisotropic contribution is the axially symmetric $l = 2$, $m = 0$ component, $S_2^0(q)$, which captures the anisotropy along the stretching direction. **Figure 11** shows $S_2^0(q)$ normalized by $S_0^0(0) = N^2$ for the ring and linear components of the M320 blend at Hencky strains $\epsilon = 0, 2, 4$, and 8.

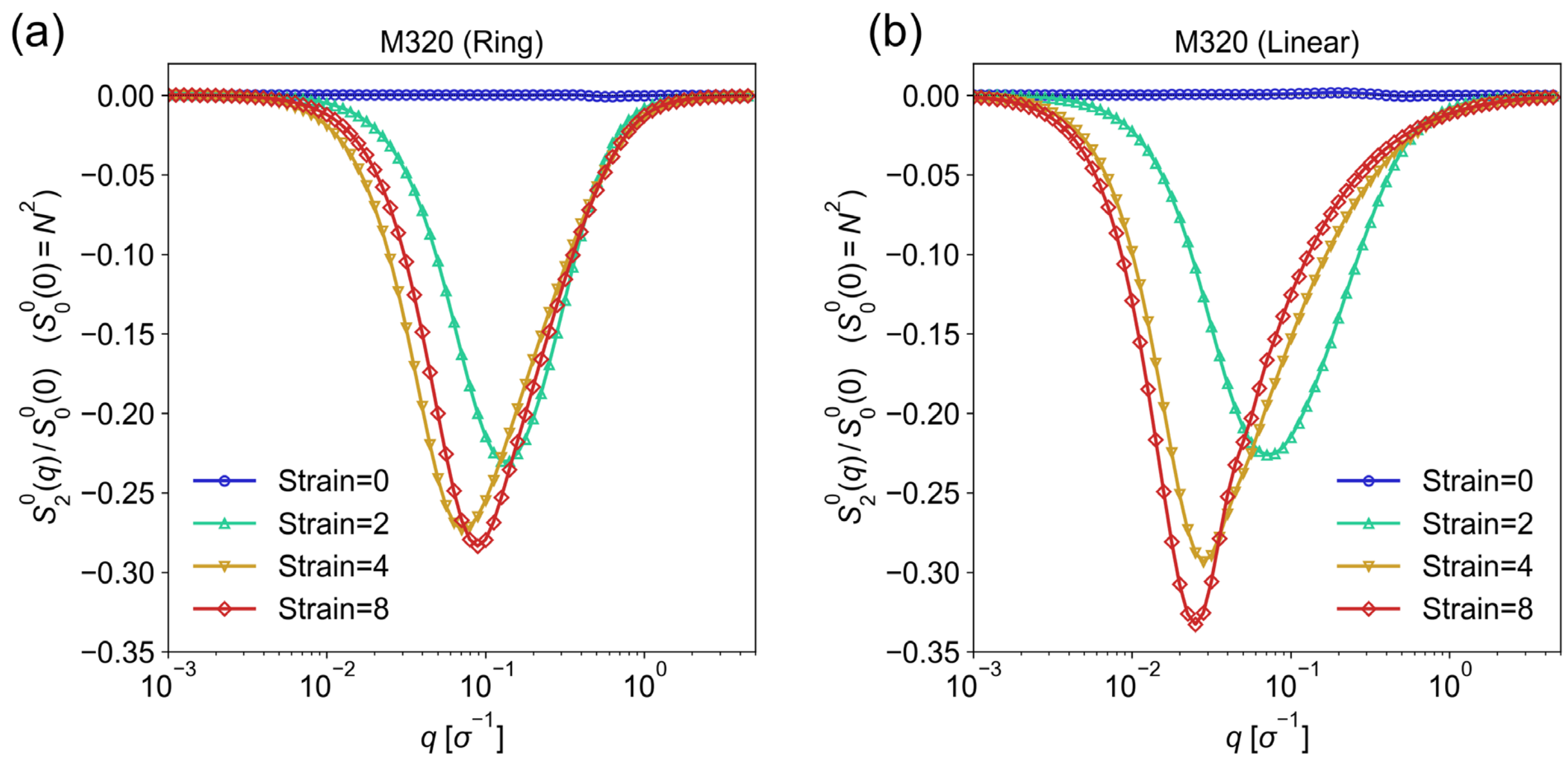

**Figure 11.** Spherical harmonic component $S_2^0(q)$ of the 2D SANS patterns, normalized by $S_0^0(0) = N^2$, for (a) the ring and (b) the linear components of the M320 blend ($\dot{\epsilon} = 0.0001$) at Hencky strains $\epsilon = 0, 2, 4$, and 8.

The normalized $S_2^0(q)$ develops a pronounced negative minimum under uniaxial elongational flow. The position of the minimum, $q_{\min}$, corresponds to the inverse of the gyration radius along the stretching direction ($q_{\min} \sim 1/R_{g,z}$): a shift of the minimum toward lower $q$ reflects elongation along the $z$ axis, while a shift toward higher $q$ reflects contraction. The depth of the minimum reflects the narrowness of the distribution of $R_{g,z}$ among chains, so that a deeper minimum corresponds to a more uniform set of chain shapes.

For the linear component (**Figure 11b**), the minimum shifts monotonically toward lower $q$ as the strain increases from 0 to 8, indicating that the linear chains elongate continuously into a rod-like conformation. For the ring component (**Figure 11a**), the minimum first shifts toward lower $q$ up to $\epsilon$ = 4 and then moves back toward higher $q$ from $\epsilon$ = 4 to 8. This non-monotonic behavior directly reflects the elongation of the rings up to the stress peak ($\epsilon$ = 4) and their subsequent recoil as the linear chains unthread ($\epsilon$ = 8), in agreement with the conformational and stress analyses in the preceding sections. In addition, the ring's minimum deepens from $\epsilon$ = 4 to 8, indicating that the ring shapes become more uniform as they relax toward the steady state. The values of $R_{g,z}$ inferred from $q_{\text{min}}$ are consistent with those obtained directly from the coordinates (**Table 2**) and from the Guinier analysis (**Table S1)**.

The successful detection of these distinct 2D SANS signatures, especially the post-peak relaxation of the ring anisotropy, would provide definitive experimental evidence for the threshold chain length and the multiple threading mechanism identified in this study. In this context, our identification of the threshold chain length at $Z \approx 4$ is of significant practical importance for experimentalists. Previous reports of stress overshoot in ring-linear blends utilized systems with $Z \approx 14$,[19,22] which require the synthesis of high-molar mass ring polymers—a task that is difficult to achieve while maintaining high topological purity. By demonstrating that the multiple threading mechanism and the resulting stress overshoot manifest at a much lower threshold ($Z \approx 4$), our work suggests that the experimental hurdle for direct structural validation is considerably lower than previously assumed. This means that researchers can focus on moderately entangled systems, for which the synthesis and LCCC purification of rings are more accessible. While reducing $Z$ eases the chemical challenges, it inherently shifts the experimental requirement toward faster deformation rates. Researchers must therefore balance the accessibility of purified rings with the technical limits of extensional rheometry, as shorter chains require significantly higher strain rates.

Nevertheless, the primary remaining challenge lies in the sophisticated material preparation required for definitive SANS characterization, specifically the synthesis of topologically pure deuterated ring polymers at sufficient quantities and purity levels for structural analysis. While pioneering SANS experiments on ring polymers[12,13,53–56] have already been reported, the synthetic constraints associated with long chains have limited the scope of direct structural validation. We hope that the lower chain length requirement identified here, combined with future breakthroughs in polymer chemistry, will soon enable the direct validation of the molecular recoil and unthreading dynamics predicted in this study.

## 5. Summary

In this study, we investigated the chain-length dependence of the uniaxial elongational rheology of equi-weight ring-linear polymer blends using coarse-grained molecular dynamics simulations. Our systematic analysis revealed a clear topological origin for the stress-overshoot phenomena, providing a unified understanding of the role of threading across different entanglement regimes. The main findings of this work are as follows:

*Identification of the Threshold Chain Length*: We identified that a threshold chain length of $Z \approx 4$ ($N = 320$) is required for the emergence of a macroscopic stress overshoot in ring-linear blends under uniaxial elongational flow. Blends with shorter chains ($Z \leq 2$) exhibit monotonic stress growth even under high strain-rate flow conditions.

*Role of Multiple Threading*: Our topological analysis using the Gauss linking number revealed that at the identified threshold ($Z \approx 4$), each ring is penetrated by multiple linear chains (averaging three to four) at equilibrium. These multiple threading events are essential to provide a sufficiently strong topological grip, allowing the rings to be stretched significantly together with the linear chains that constitute the transient entanglement network.

*Molecular Mechanism of Overshoot*: The stress overshoot is quantitatively linked to a transient thread-to-unthread transition. Stress decomposition and conformational analysis confirmed that the peak stress corresponds to the maximum extension of the rings, followed by a stress drop as the linear chains unthread, leading to the structural recoil of the ring component.

*Structural Evolution via 2D Scattering Patterns*: We demonstrated that the 2D scattering patterns of the single ring component exhibit a distinct evolution that reflects the macroscopic stress response. As the rings reach their maximum extension at the stress peak, the scattering intensity becomes sharply concentrated at $q_z \approx 0$ and expanded along the axis perpendicular to the elongation direction ($q_\perp$). Subsequently, this intense anisotropy diminishes as the rings undergo structural recoil during the unthreading process, providing a direct experimental signature for validating the predicted molecular recoil and unthreading dynamics.

We note that the identified threshold of $Z \approx 4$ should be understood within the scope of the present model and conditions. Our simulations were performed for fully flexible Kremer–Grest chains at a fixed 1:1 ring-linear composition. Expressing the threshold in terms of the number of entanglements $Z = N/N_e$ partly absorbs the dependence on the entanglement length $N_e$. The emergence of a macroscopic overshoot in the total stress, however, depends not only on the overshoot of the ring component but also on how rapidly the linear component grows under flow, and this balance may be shifted by chain stiffness and blend composition. A systematic study of these dependencies is therefore left for future work.

Looking forward, our finding that the stress overshoot manifests at a relatively low entanglement number ($Z \approx 4$) provides an important guideline for future experimental validations. While previous studies[19,22] focused on high-molar mass systems ($Z \approx 14$), which are notoriously difficult to synthesize and purify, our results suggest that moderately entangled systems are sufficient to capture the physics of the thread-to-unthread transition. Indeed, comprehensive SANS (and neutron spin echo) investigations by Kruteva et al.[12,13] have already established that the fundamental conformational and dynamic signatures of ring-linear topological interactions—such as threading-induced swelling and local relaxation modes—are clearly detectable in pure and blended systems. However, the high synthetic effort required to produce high-purity deuterated rings at very large $Z$ has limited the exploration of non-linear structural evolution under fast flows. In the moderately entangled regime ($Z \approx 4$), the preparation of topologically pure samples via LCCC is more accessible. We hope that the predicted 2D SANS patterns for $Z \approx 4$ systems will encourage experimentalists to utilize moderately entangled, deuterated ring polymers to directly observe the molecular recoil and unthreading dynamics predicted in this study.

ASSOCIATED CONTENT

**Supporting Information**.

The following file is available free of charge.

SupportingInformation.pdf (PDF)

Robustness of the first normal stress difference (smoothing, system size, and initial configuration); stress overshoot behavior under uniaxial elongational flow at a longer chain system; rheological decomposition of shorter chain blends; 1D SANS profiles and Guinier plots.

AUTHOR INFORMATION

**Corresponding Author**

**Takahiro Murashima** – Department of Physics, Tohoku University, 6-3, Aramaki-aza-Aoba, Aoba-ku, Sendai, 980-8578, Japan; Email: murasima@cmpt.phys.tohoku.ac.jp

**Authors**

**Katsumi Hagita** – Department of Applied Physics, National Defense Academy, 1-10-20 Hashirimizu, Yokosuka, 239-8686, Japan.

**Toshihiro Kawakatsu** – Admissions Center, Tohoku University, 28, Kawauchi, Aoba-ku, Sendai, 980-8576, Japan.

**Author Contributions**

T.M. led the study, performed the coarse-grained molecular dynamics simulations, conducted the data analysis, and developed the physical interpretations. K.H. and T.K. provided technical support and participated in the discussion of the results. All authors have reviewed and approved the final version of the manuscript

**Funding Sources**

This work was partly financially supported by JSPS KAKENHI, Japan (grant Nos. JP18H04494, JP19H00905, JP20K03875, JP20H04649, JP21H00111, JP25K01839, and JP25K07238), JST

CREST, Japan, (JPMJCR1993, JPMJCR19T4, and JPMJCR25Q3), and the Fuji Seal Foundation (T.M.).

ACKNOWLEDGMENT

T.M. thanks Prof. T. Taniguchi, Prof. M. Sugimoto, Prof. J.-I. Takimoto, Prof. S. K. Sukumaran, Prof. T. Uneyama, Prof. Y. Masubuchi, Prof. Y. Doi, Dr. T. Honda, Dr. Y. Tomiyoshi, Dr. T. Sato, and Dr. N. Sakata for their fruitful discussions, comments, and encouragement. The authors thank Prof. H. Jinnai, Prof. T. Satoh, Prof. T. Deguchi, and Prof. K. Shimokawa for their support and encouragement. T.M. and T.K. thank the collaboration program of the Advanced Imaging and Modeling Center for Soft-materials (AIMcS) of Tohoku University. For the computations in this work, the authors were partially supported by the Supercomputer Center, Institute for Solid State Physics, University of Tokyo; MASAMUNE-IMR at the Center for Computational Materials Science, Institute for Materials Research, Tohoku University; Grand Chariot, Grand Chariot 2, and Polaire at the Hokkaido University Information Initiative Center; Flow at Nagoya University Information Technology Center; SQUID at Cybermedia Center, Osaka University; Genkai at Research Institute for Information Technology, Kyushu University; Miyabi at Information Technology Center, University of Tokyo; Fugaku at RIKEN Center for Computational Science; the Joint Usage/Research Center for Interdisciplinary Large-Scale Information Infrastructures (JHPCN); and the High-Performance Computing Infrastructure (HPCI) in Japan (hp200048, hp200168, hp210102, hp210132, hp220019, hp220104, hp220113, hp220114, hp230384, hp240469, hp250329, hp250144, hp260115, jh210035, jh220038, jh230026, jh250040, and jh260052). This work was partly financially supported by JSPS KAKENHI, Japan (grant Nos. JP18H04494, JP19H00905, JP20K03875, JP20H04649, JP21H00111, JP25K01839, and JP25K07238), JST CREST, Japan, (JPMJCR1993,

JPMJCR19T4, and JPMJCR25Q3), and the Fuji Seal Foundation (T.M.). The authors thank Editage (www.editage.com) for the English language editing.

# Supporting Information for "*Identifying the Threshold Chain Length for Stress Overshoot in Ring/Linear Polymer Blends under Uniaxial Elongation: The Role of Multiple Threading*"

*Takahiro Murashima,*[1] Katsumi Hagita,[2] and Toshihiro Kawakatsu[3]*

[1]Department of Physics, Tohoku University, 6-3, Aramaki-aza-Aoba, Aoba-ku, Sendai, 980-8578, Japan

[2]Department of Applied Physics, National Defense Academy, 1-10-20 Hashirimizu, Yokosuka, 239-8686, Japan

[3]Addmissions Center, Tohoku University, 28, Kawauchi, Aoba-ku, Sendai, 980-8576, Japan

*murasima@cmpt.phys.tohoku.ac.jp

## S1. Robustness of the First Normal Stress Difference (smoothing, system size, and initial configuration)

In this section, we demonstrate that the first normal stress difference $N_1 = \sigma_{zz} - 0.5(\sigma_{xx} + \sigma_{yy})$ reported in the main text is robust with respect to three factors: the data smoothing, the system size, and the initial configuration.

The first normal stress difference $N_1$ was smoothed using a first-order Savitzky-Golay filter [1] on a logarithmic time grid (20 points per decade). The half-window width was chosen adaptively, widening from 2 input points at short times to at most 1000 at long times, so that the broader window suppresses the larger noise of the raw data at longer times. **Figure S1** compares the raw and smoothed $N_1$ as a function of strain $\epsilon = t\dot{\epsilon}$ at high, intermediate, and low strain rates ($\dot{\epsilon} = 10^{-3}$, $10^{-4}$, and $10^{-5}$), for both the system size used in the main text (614,400 particles, top row) and a system eightfold larger (4,915,200 particles, bottom row). The noise in the raw data, which is most pronounced at the lowest strain rate, is markedly reduced in the larger system; nevertheless, the overshoot peak persists in both system sizes. At the highest strain rate ($\dot{\epsilon} = 10^{-3}$), a slight discrepancy between the raw and smoothed curves is visible, which could be mitigated by narrowing the half-window width at short times; however, since the focus of this study is on confirming the overshoot rather than capturing its precise shape, we adopted the same smoothing protocol across all strain rates for consistency. This confirms that the overshoot—including its height and position—is a genuine physical feature rather than an artifact of the smoothing or of finite system size.

To assess the dependence of the results on the initial configuration, we compared $N_1$ as a function of strain for five deformation runs (main and init1 to init4) using equilibrium configurations taken from a single equilibrium trajectory as initial states. Successive configurations were separated by an equilibration interval of $10^6$ τ, which is longer than the longest relaxation time of the system, so that they can be regarded as effectively independent. **Figure S2** shows this comparison for the full system and for the ring and linear components separately, at high, intermediate, and low strain rates ($\dot{\epsilon} = 10^{-3}$, $10^{-4}$, and $10^{-5}$). The curves from different initial configurations agree closely across all components and shear rates. The small run-to-run scatter appears mainly in the height of the overshoot peak, while the peak position and the steady-state value are essentially unaffected, confirming that the reported nonlinear behavior is robust with respect to the initial configuration.

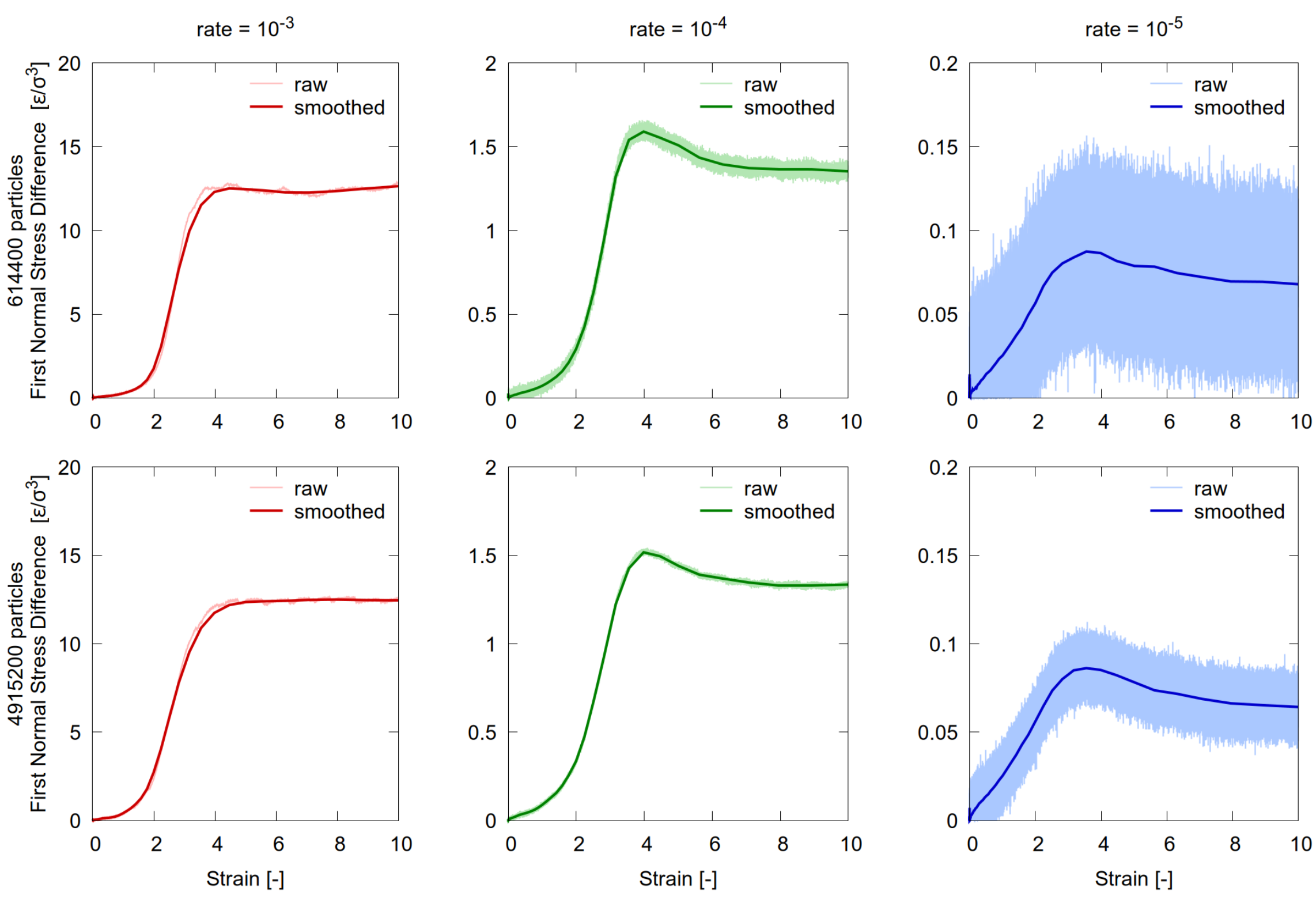

**Figure S1.** Raw (light) and smoothed (dark) first normal stress difference $N_1 = \sigma_{zz} - 0.5(\sigma_{xx} + \sigma_{yy})$ as a function of strain for three shear rates, $\dot{\epsilon} = 10^{-3}$ (left, red), $10^{-4}$ (middle, green), and $10^{-5}$ (right, blue). The top row shows the system size used in the main text (614,400 particles) and the bottom row a system eightfold larger (4,915,200 particles).

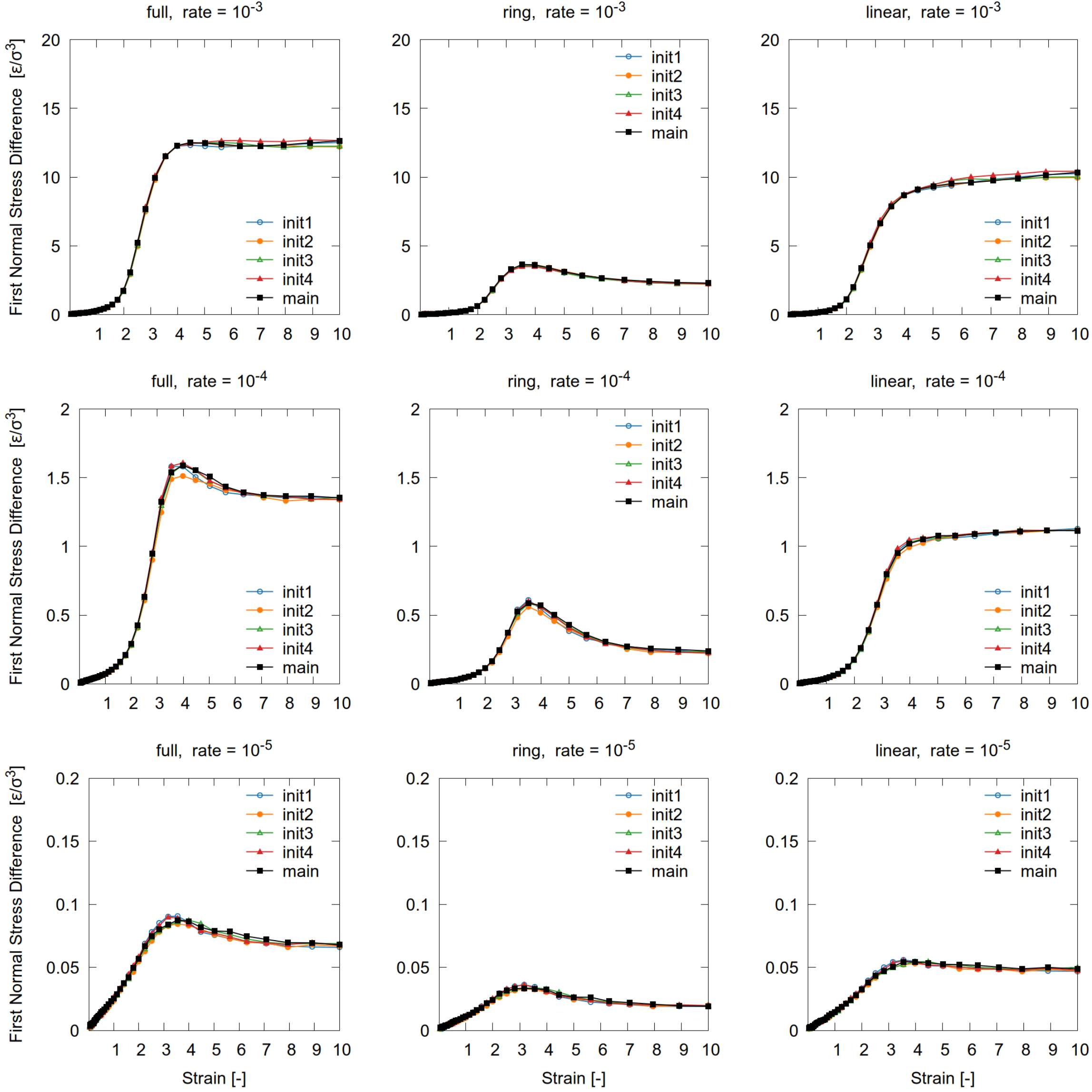

**Figure S2.** First normal stress difference $N_1$ as a function of strain for four independent runs from distinct initial configurations (init1-init4) and the run used in the main text (main), shown for the full system (left), the ring component (middle), and the linear component (right), at shear rates $\dot{\epsilon} = 10^{-3}$ (top), $10^{-4}$ (middle), and $10^{-5}$ (bottom).

## S2. Stress Overshoot Behavior under Uniaxial Elongational Flow at a Longer Chain System

To further validate the chain-length dependence of the stress overshoot, we performed additional simulations for a 1:1 ring-linear blend of flexible chains with $N = 640$ ($Z = 8$). The simulation methodology and parameters are identical to those described in the Main Text. Figure S3 shows the transient uniaxial elongational viscosity $\eta_E(t)$ for this longer chain system. Compared to the $Z = 4$ case discussed in the Main Text, the $Z = 8$ system exhibits a more pronounced and sharper stress overshoot at equivalent strain rates. This enhancement is consistent with our hypothesis that an increased number of topological constraints—specifically, a higher degree of threading—leads to greater conformational tension within the rings.

**Figure S3.** Transient uniaxial elongational viscosity $\eta_{\mathrm{E}}(t) = \{\sigma_{zz}(t) - 0.5[\sigma_{xx}(t) + \sigma_{yy}(t)]\} / \dot{\epsilon}$ $[\boldsymbol{\tau\varepsilon/\sigma^3}]$ as a function of time $t$ $[\boldsymbol{\tau}]$ for a 1:1 ring-linear blend of flexible chains with $N = 640$ ($Z = 8$) under uniaxial elongational flow. The numerical values denote the applied strain rates, $\dot{\epsilon}$ $[1/\tau]$. All physical quantities are expressed in Lennard-Jones (LJ) units.

## S3. Rheological Decomposition of Shorter Chain Blends

To reinforce the identification of the threshold chain length ($Z \approx 4$) discussed in the Main Text, we provide the component-wise stress decomposition for the shorter chain blends, namely, M160 and M80.

**Figure S4** shows the transient first normal stress difference $N_1 = \sigma_{zz} - 0.5(\sigma_{xx} + \sigma_{yy})$ for the M160 blend ($N$ =160, $Z$ = 2). While the ring component (**Figure S4b**) exhibits a slight stress overshoot at the higher strain rates, its magnitude is insufficient to induce a clear overshoot in the total stress of the blend (**Figure S4a**). In the total stress response, monotonic growth remains dominant, reflecting a state where the degree of threading is insufficient to sustain high conformational tension across the network.

**Figure S5** shows the results for the M80 blend ($N$ =80, $Z$ = 1). At this chain length, which is near or below the entanglement/threading threshold, no stress overshoot is observed in either the ring or the linear component. Both species exhibit purely monotonic growth toward a steady state (**Figure S5b and c**), confirming that the "thread-to-unthread" transition requires a minimum level of topological constraint that is absent in such short-chain systems.

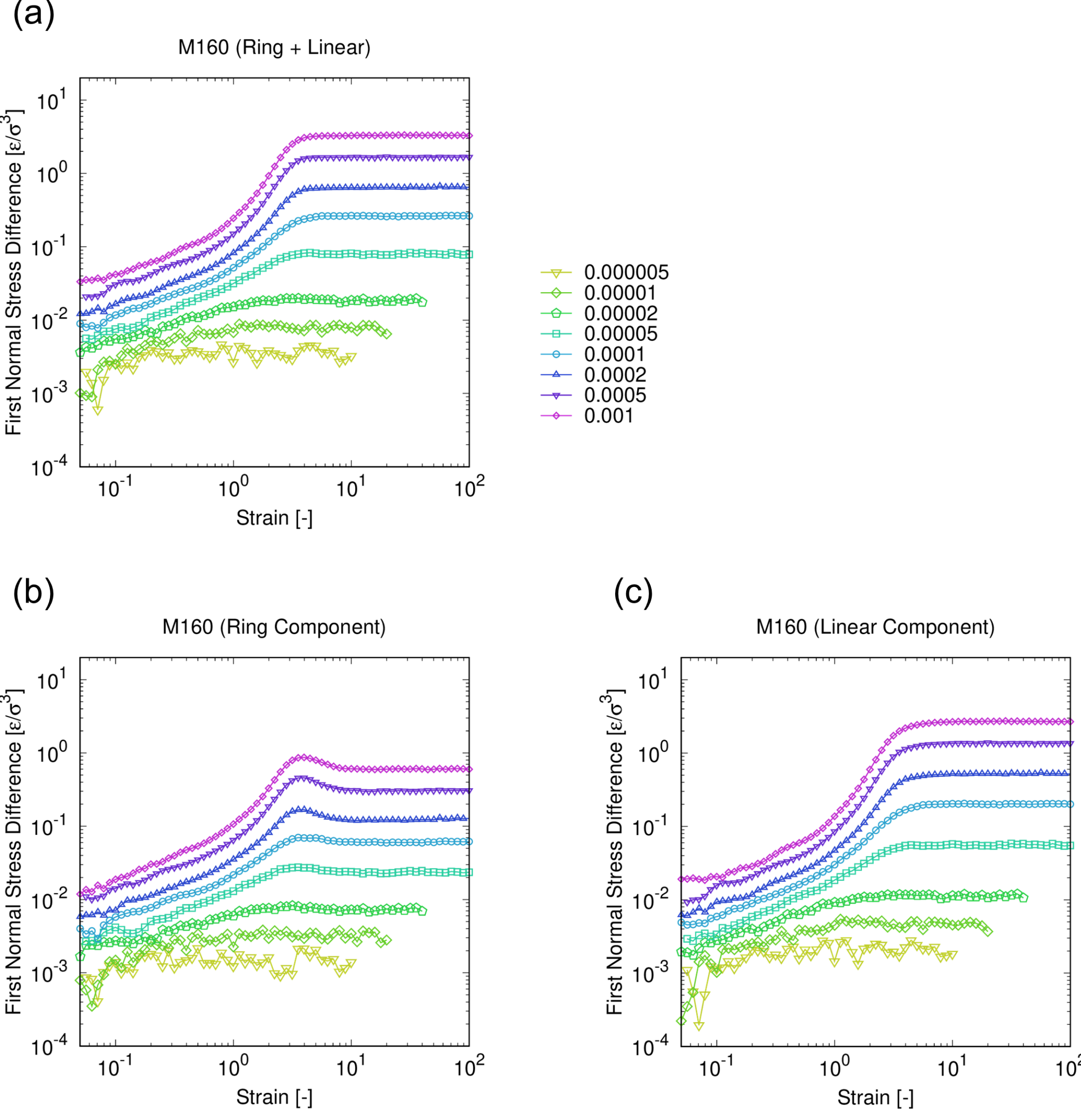

**Figure S4.** Transient first normal stress difference $N_1 = \sigma_{zz} - 0.5(\sigma_{xx} + \sigma_{yy})$ as a function of Hencky strain for the M160 blend ($N$ =160, $Z$ = 2) at various strain rates. (a) Total stress of the blend, (b) contribution from the ring component, and (c) contribution from the linear component. While the ring component (b) exhibits a slight stress overshoot at high strain rates, its magnitude is insufficient to induce a clear overshoot in the total stress (a) of the blend, where monotonic growth remains dominant. This contrast with the M320 case highlights the necessity of a higher degree of threading for macroscopic non-linear rheological signatures.

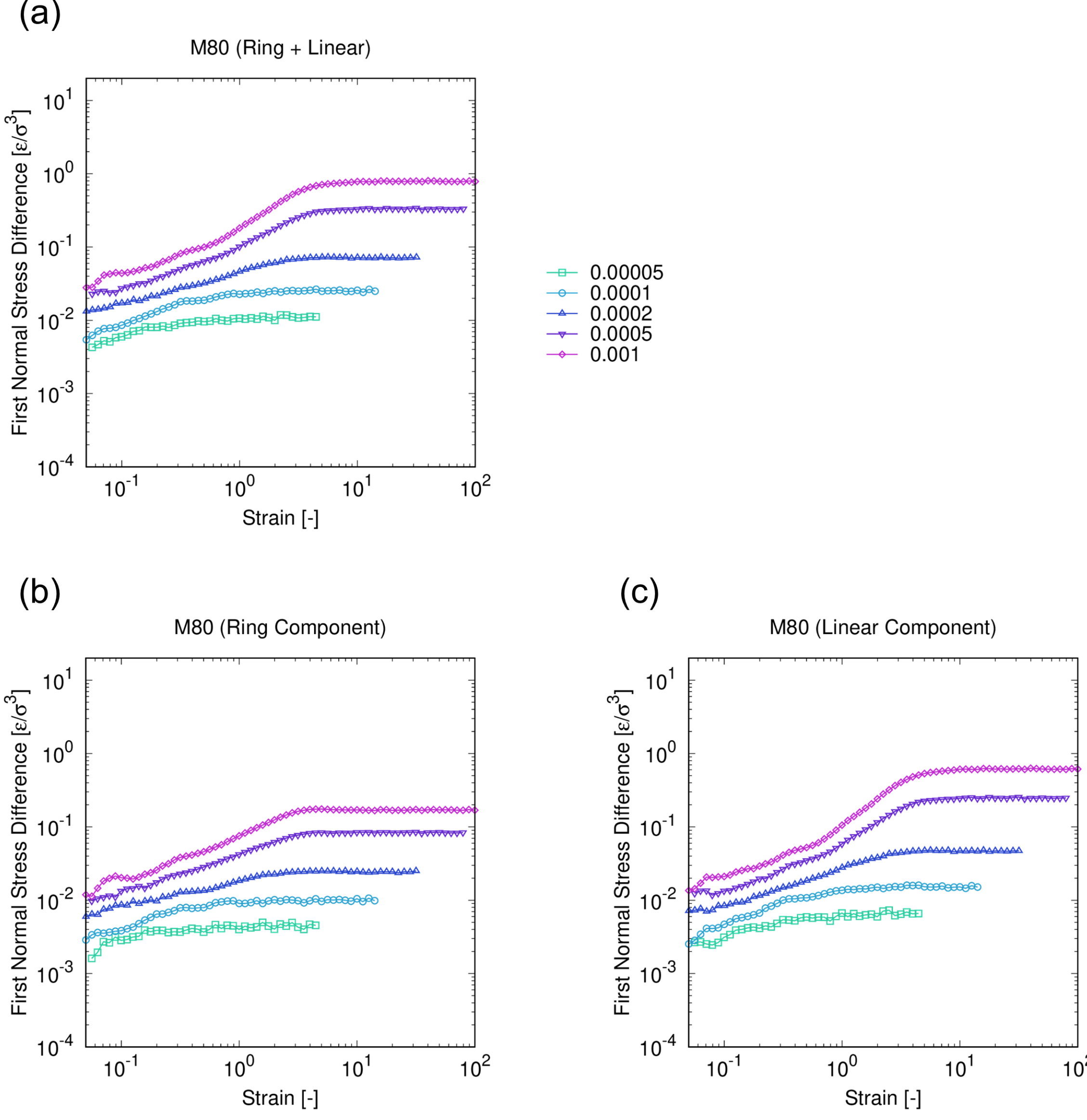

**Figure S5.** Transient first normal stress difference $N_1 = \sigma_{zz} - 0.5(\sigma_{xx} + \sigma_{yy})$ as a function of Hencky strain for the M80 blend ($N$ =80, $Z$ = 1) at various strain rates. (a) Total stress of the blend, (b) contribution from the ring component, and (c) contribution from the linear component. At this chain length (below the entanglement/threading threshold), the stress grows monotonically for both components.

## S4. 1D SANS Profiles and Guinier Plots

To quantitatively evaluate the structural changes under strain, 1D scattering profiles were extracted from the 2D-SANS patterns (**Figure S6**). The profiles along the $q_\perp$ direction were obtained by taking a slice at $q_z = 0$, while the profiles along the $q_z$ direction were obtained at $q_\perp = 0$. In both ring and linear components in the blend of M320, the $q_z$ profiles exhibit a rapid decay with increasing strain, indicating structural elongation along the *z*-axis.

Guinier analysis was performed on these 1D profiles in the low-$q$ region to determine the components of the radius of gyration. **Figure S7** shows the Guinier plots ($\ln[I(q)/I(0)]$ vs. $q^2$) for both directions. The linear fits of these plots yielded the values of $R_{g,z}$ and $R_{g,\perp}$ as summarized in the legends of **Figure S7** and **Table S1**. It should be noted that for the 1D profiles extracted along the principal axes, the Guinier approximation is expressed as: $\ln I(q_\alpha) \approx \ln I(0) - R_{g,\alpha}^2 q_\alpha^2$ $(\alpha = z, \perp)$. In contrast to the conventional 3D Guinier law where a factor of 1/3 is applied to the total $R_g^2$, the factor for these 1D line-cut profiles is unity. This is because the profiles represent the scattering along a specific direction, allowing for the direct determination of the directional components of the radius of gyration, $R_{g,z}$ and $R_{g,\perp}$, from the slopes of the Guinier plots.

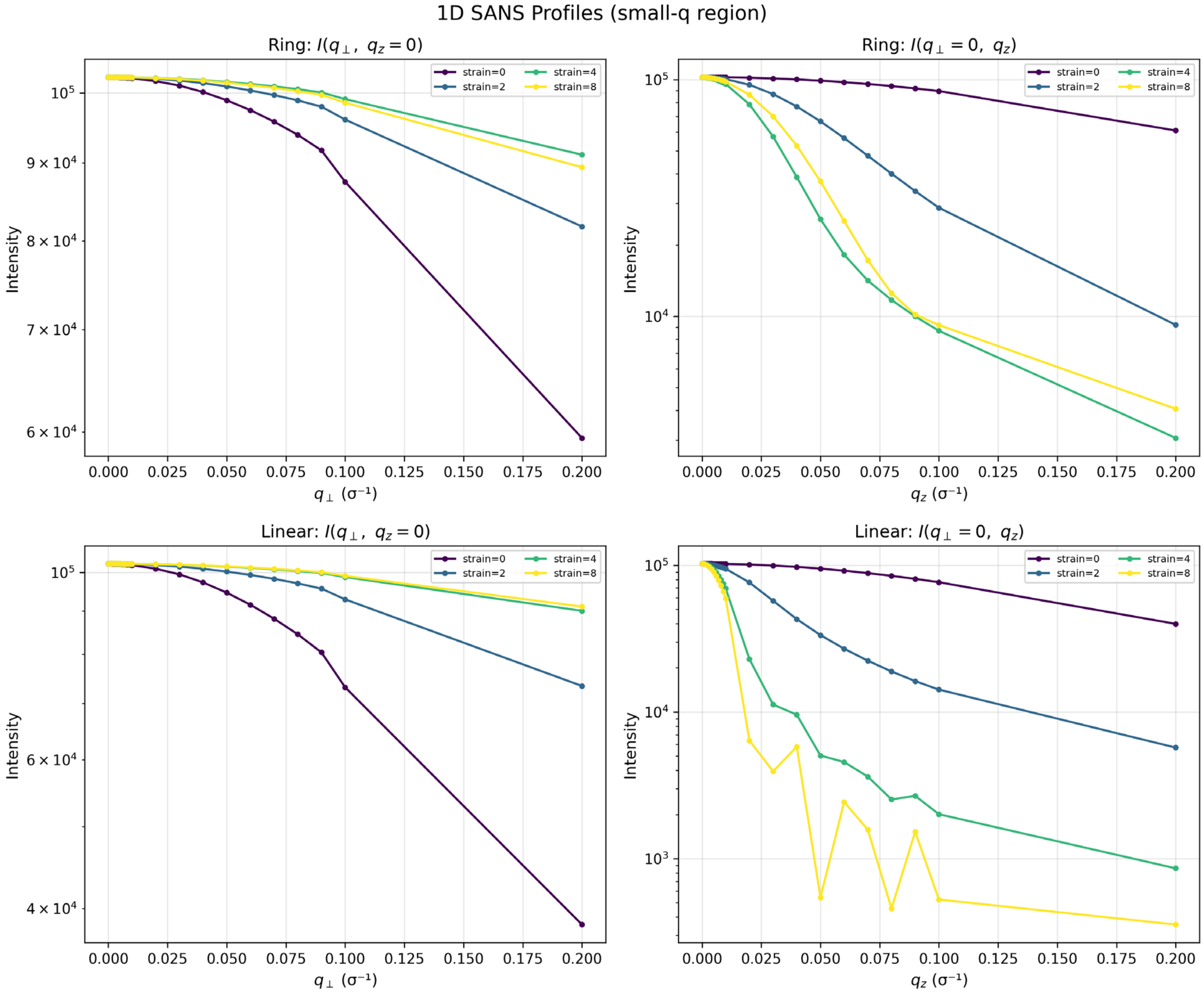

**Figure S6.** 1D SANS intensity profiles in the low-$q$ region. The profiles were extracted from the 2D-SANS patterns for (top) ring and (bottom) linear components in the blend of M320 at various strains ($\epsilon$ = 0, 2, 4, and 8). (Left) $I(q_\perp, q_z = 0)$ profiles along the perpendicular direction and (right) $I(q_\perp = 0, q_z)$ profiles along the parallel (elongation) direction.

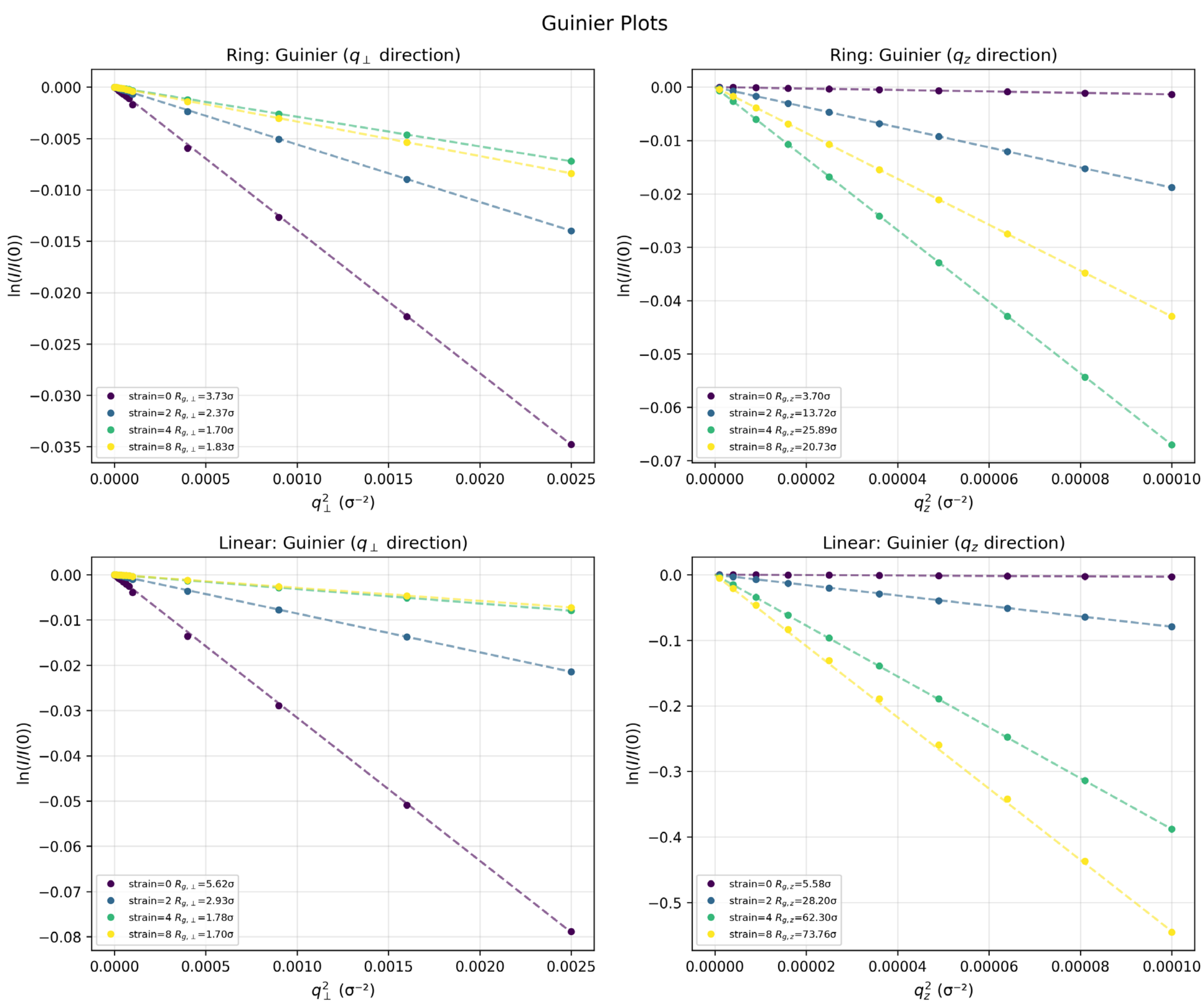

**Figure S7.** Guinier plots extracted from the 1D SANS profiles. $\ln[I(q)/I(0)]$ is plotted against $q^2$ for (top) ring and (bottom) linear components in the blend of M320. (Left) Plots for the $q_\perp$ direction and (right) plots for the $q_z$ direction. The dashed lines represent the linear fits used to determine the directional components of the radius of gyration, $R_{g,\perp}$ and $R_{g,z}$.

**Table S1.** Radius of gyration components along z-axis ($R_{g,z}$) and perpendicular direction ($R_{g,\perp}$) for rings and linear chains in the blend of M320 determined by Guinier analysis. Note that the radius of gyration $R_g$ is estimated by $R_g^2 = R_{g,z}^2 + 2R_{g,\perp}^2$.

| | Topology | Strain = 0 | Strain = 2 | Strain = 4 | Strain = 8 |
|---|---|---|---|---|---|
| $R_g$ | Ring | $6.4\sigma$ | $14.1\sigma$ | $26.0\sigma$ | $20.8\sigma$ |
| | Linear | $9.7\sigma$ | $28.5\sigma$ | $62.3\sigma$ | $73.7\sigma$ |
| $R_{g,z}$ | Ring | $3.6\sigma$ | $13.7\sigma$ | $25.8\sigma$ | $20.7\sigma$ |
| | Linear | $5.5\sigma$ | $28.1\sigma$ | $62.3\sigma$ | $73.7\sigma$ |
| $R_{g,\perp}$ | Ring | $3.7\sigma$ | $2.3\sigma$ | $1.7\sigma$ | $1.8\sigma$ |
| | Linear | $5.6\sigma$ | $2.9\sigma$ | $1.7\sigma$ | $1.7\sigma$ |